\documentstyle[10pt,aaspp4,psfig]{article}
\font\cap=cmcsc10

\def\ni{\noindent}        

\def\hi{\noindent \hangindent=2.5em}

\def\kpc{{\rm\,kpc}}
\def\Mpc{{\rm\,Mpc}}

\def\hnot{{\rm\,km/s/Mpc}}

\def\surfb{{\rm\,mag/arcsec^2}}

\def\araa{{Ann.\ Rev.\ Astr.\ Ap.}, }
\def\aj{{A.~J.}, }  
\def\apj{{Ap.~J.}, }  
\def\apjs{{Ap.~J.~Suppl.}, }  
\def\apjl{{Ap.~J.~(Letters)}, } 
\def\mn{{M.N.R.A.S.}, }      
\def\nat{{Nature}, }      
\def\aa{{Astr.~Ap.}, }     
\def\aasup{{Astr.~Ap.~Suppl.}, }     

\def\spose#1{\hbox to 0pt{#1\hss}}
\def\lta{\mathrel{\spose{\lower 3pt\hbox{$\mathchar"218$}}
     \raise 2.0pt\hbox{$\mathchar"13C$}}}
\def\gta{\mathrel{\spose{\lower 3pt\hbox{$\mathchar"218$}}
     \raise 2.0pt\hbox{$\mathchar"13E$}}}

\def\clock{\count0=\time \divide\count0 by 60
     \count1=\count0 \multiply\count1 by -60 \advance\count1 by \time
     \number\count0:\ifnum\count1<10{0\number\count1}\else\number\count1\fi}

\begin{document}

\title{Systematic Biases in Galaxy Luminosity Functions}

\author{Julianne J. Dalcanton\altaffilmark{1,2}}
\affil{Observatories of the Carnegie Institution
	of Washington, 813 Santa Barbara Street, Pasadena CA, 91101}
\bigskip
\centerline{\it Accepted to the Astrophysical Journal}
\medskip

\altaffiltext{1}{e-mail address: jd@ociw.edu}
\altaffiltext{2}{Hubble Fellow}
  
\begin{abstract}

Both the detection of galaxies and the derivation of the luminosity
function depend upon isophotal magnitudes, implicitly in the first
case, and explicitly in the latter.  However, unlike perfect point
sources, the fraction of a galaxy's light contained within the
limiting isophote is a function of redshift, due to the combined
effects of the point spread function and cosmological dimming.  This
redshift variation in the measured isophotal luminosity can strongly
affect the derived luminosity function.  Using simulations which
include the effects of seeing upon both disk and elliptical galaxies,
we explore the size of the systematic biases which can result from
ignoring the redshift variation in the fraction of detected light.  We
show that the biases lead to underestimates in the normalization of
the luminosity function, as well as changes in shape.  The size of the
bias depends upon redshift, and thus can mimic galaxy evolution.
Surprisingly, these biases can be extremely large {\it without}
affecting $\langle V/V_{max} \rangle$.  However, these biases can be
detected in the full distribution of $V/V_{max}$, and in fact may have
already been detected in recent surveys.  Because the systematic
biases result from the redshift variation in the fraction of lost
light, the biases are not significant when the fraction of lost light
is always small over the entire survey volume, for all galaxy types.
However, as modern galaxy surveys now reach higher redshifts, lower
surface brightnesses, and smaller angular sizes, the effects of seeing
and galaxy visibility are becoming increasingly important and need to
be taken into account.  We show that the expected biases are not
necessarily eliminated when using aperture magnitudes, FOCAS ``total''
magnitudes, or Kron magnitudes, but may be reduced significantly if
Petrosian magnitudes are used.  These considerations may also apply to
samples of clusters selected in X-rays.

\end{abstract}

\section{Introduction}

The luminosity function remains one of the principal tools for
quantifying the population of galaxies.  The evolution of the luminosity
function with redshift provides a critical test for assessing changes
in the galaxy population with time.  Variations in the galaxy
luminosity function with environment, emission line strength, and
morphological type all provide broad clues towards distinguishing
among galaxy formation scenarios.  Given that the usefulness of the
luminosity function as a cosmological tool depends upon comparing
luminosity functions at different times or for different galaxy types,
it is extremely important that the measurement of the luminosity
function be unbiased with redshift, galaxy morphology, or survey
technique.

Standard approaches for calculating the luminosity function involve
mathematical variants of what Binggeli et al.\ (1988) refer to as the
``classical method'', wherein each galaxy is weighted by the inverse
of the maximum volume over which it could have been detected, given
the galaxy's luminosity and the magnitude limit of the survey.
Other methods which reduce the impact of spatial inhomogeneity on the
derived luminosity function, such as Sandage et al.'s (1979)
parametric maximum-likelihood method, calculate a likelihood based
upon comparing the observed luminosity of each galaxy in the
sample to the range of potentially observable luminosities at
each galaxy's redshift.

Both of these methods require estimating the range of absolute
magnitudes which can be seen at each redshift, given the selection
criteria of a survey.  Any systematic error in this procedure can lead
to a systematic mismeasurement of the luminosity function.  As we will
show here, typical procedures for measuring the luminosity function
do indeed suffer from these systematic biases, largely due to a
simplistic estimate of how a galaxy's apparent magnitude varies with
redshift.

Most determinations of the luminosity function assume that an
isophotal magnitude is a good representation of the galaxy's total
magnitude, and that the isophotal magnitude varies with redshift as
the inverse square of the luminosity distance (modulo
$k$-corrections).  While these are perhaps reasonable approximations
for point sources in deep data with exceptional seeing, or for nearby
surveys of extremely bright, high surface brightness galaxies, these
assumptions break down for real galaxies in many modern surveys
(e.g. McGaugh 1994).  Some of a galaxy's light is lost beyond the
outer isophote of a galaxy image, due to the combined effects of
seeing and surface brightness variations (both cosmological and
intrinsic), leading to the apparent isophotal magnitude falling off
more rapidly with redshift than predicted.  Thus, field galaxy surveys
can easily overestimate the effective volume of the survey, and
therefore systematically underestimate the luminosity function.  More
importantly, the degree to which the derived luminosity function is
underestimated is a strong function of galaxy size and surface
brightness, as well as being redshift dependent.  This can lead to
apparent redshift evolution in the galaxy population, and to erroneous
conclusions about the variation of the morphological mix within the
galaxy population.  The measured luminosity function can also depend
upon seeing such that similar surveys with different imaging data
(e.g. ground based vs.\ HST) can produce different results.  While
these biases can be mitigated against by using exceptionally deep
imaging data or large aperture magnitudes, the majority of deep field
surveys have used isophotal magnitudes to derive the luminosity
function.  Instead of isophotal magnitudes, some fraction of surveys
have used corrected ``total'' magnitudes which choose an aperture size
dynamically, based upon either the isophotal area or the
intensity-weighted first moment radius.  However, in the former case,
the corrected ``total'' magnitude actually reduces to using an
isophotal magnitude with a fainter limiting magnitude, as we will show
in \ref{petrosiansect}.  Furthermore, even when isophotal magnitudes
are not used for photometry, the selection of galaxies for a survey
depends implicitly on the isophotal magnitude limit of the survey
images, subjecting the derived luminosity function to many of the
biases we explore here.

The importance of correcting for light lost beyond the limiting
isophote has been long known (Humason et al.\ 1956), and has
traditionally been studied through curve-of-growth analyses.  Sandage
(1961) explicitly warns against the use of isophotal magnitudes in
cosmological tests, pointing out that they are not a consistent
measure of galaxy flux at all redshifts.  Until the recent era of deep
redshift surveys, it has been sufficient to use growth curves to
correct isophotal magnitudes onto the same magnitude system; since
surveys covered only a limited range in redshift, once galaxy
magnitudes were corrected to the same isophotal limit, a constant
fraction of light was detected for each morphological type over the
entire redshift range spanned by the data.  In contrast, redshift
surveys now routinely reach redshifts of 1 (e.g.\ CFRS, AUTOFIB; Lilly
et al.\ 1995a, Ellis et al.\ 1996), such that even with a single
limiting isophote, the fraction of detected light for a galaxy is not
the same at all redshifts.  Furthermore, deep redshift surveys
typically have much deeper limiting isophotes for both detection and
photometry than do the local redshift surveys to which they compare,
and typically do not make a correction onto a common isophotal
magnitude limit.  This discrepancy can lead to an apparent increase in
the number density of galaxies with increasing redshift, as pointed
out by McGaugh (1994), and explored in Ferguson \& McGaugh (1995) and
Phillipps \& Driver (1995).  Yoshii (1993) used similar techniques as
will be outlined in this paper to calculate how the use of isophotal
magnitudes affect the number counts and redshift distributions of
faint galaxies.

In this paper, we investigate how biases resulting from the use of
isophotal magnitudes affects the luminosity function in particular.
We attempt to provide a guide for estimating the degree to which these
biases may be present in different galaxy surveys.  First, in
\S\ref{Vmaxsect}, we outline the method for calculating the true
volume of a galaxy survey using the fraction of observed light as a
function of redshift, which we calculate in \S\ref{fsect} as a
function of observational parameters and intrinsic galaxy properties.
We quantify the degree to which this will produce misestimates of the
luminosity function, such that one can identify the observational
regimes where these effects are not important.  In \S\ref{lumfuncsect}
we use artificial galaxy catalogs to demonstrate how the measured
luminosity function varies when different reconstructions of the
luminosity function are used.  In \S\ref{Vsect}, we discuss ways in
which the distribution of $V/V_{max}$ can be used to test for the
presence of luminosity function biases and in \S\ref{petrosiansect},
we discuss the merits and drawbacks of alternatives to
isophotal magnitudes.

\section{$V_{max}$ for Extended Objects}	\label{Vmaxsect}

Measurements of the field luminosity function begin by calculating an
absolute magnitude for each galaxy in a redshift survey, using

\begin{equation}				\label{Morigeqn}
M = m - 25 - 5\log{\left(\frac{D_L(z)}{\Mpc}\right)} - 2.5\log{k(z)},
\end{equation}

\ni where $m$ is the apparent magnitude of an observed galaxy at
redshift $z$, $D_L(z)$ is the luminosity distance appropriate for the
chosen cosmology and $k(z)$ is the $k$-correction, which incorporates
changes in magnitude due to the stretching of the bandpass and the
variation in the observed rest-frame wavelength with redshift.

The relationship between apparent magnitude and absolute magnitude
given in equation \ref{Morigeqn} is only correct in the limit where
the apparent magnitude $m$ is an accurate measure of the total flux,
regardless of redshift or morphological type.  More typically,
however, the apparent magnitude which is used is either an isophotal
magnitude (e.g.\ Lilly et al.\ 1995b, Ellis et al.\ 1996, Lin et al.\
1996), a ``total'' magnitude measured within some multiple of the
isophotal area (Small et al.\ 1997), a modified Kron (1980) magnitude
measured within an aperture whose size is determined by the first
moment radius of the light visible above some limiting isophote (Yee
et al.\ 1996, Lin et al.\ 1997) or, more rarely, an aperture magnitude
(Gardner et al.\ 1996, Glazebrook et al.\ 1994); we will restrict
ourselves to the case of isophotal magnitudes for the purposes of this
paper, since isophotal magnitudes are most commonly used, and other
magnitude measures reduce to this case; we will consider alternative
measures of magnitude in \S\ref{petrosiansect}.

For an isophotal magnitude measured within a limiting isophote
$\mu_{lim}$, some fraction of the light is lost outside the outer
isophote.  The fraction of detected light, $f(z)$, depends both upon
intrinsic properties of the galaxy (such as the intrinsic central
surface brightness $\mu_0$, the true absolute magnitude $M$, the
two-dimensional shape of the galaxy in the absence of seeing (i.e.\
its light profile)), as well as observational parameters of the
survey, such as the limiting isophote $\mu_{lim}$ and the point spread
function.

The fraction of detected light also depends upon the redshift of the
observed galaxy.  As a galaxy moves to higher redshifts, it suffers
two effects which rapidly decreases the fraction of light
detected within a fixed isophotal limit.  First, at large enough redshifts
the galaxy appears small compared to the point spread function, and
begins to lose light beyond the limiting isophote, due to the rapid
falloff in the point spread function with radius.  Second, the
apparent surface brightness drops off as $(1+z)^{-4}$ due to the
difference in the redshift dependences of the angular diameter and
luminosity distances.  As the drop in apparent surface brightness
becomes significant (a factor of 2 at $z=0.2$), a larger fraction of a
galaxy's light falls below the limiting isophote, again increasing the
fraction of lost light.  The direction of both of these effects is for
the apparent magnitude to drop off more quickly with distance than
predicted by equation \ref{Morigeqn}.

If instead, one includes the correction for lost light, equation
\ref{Morigeqn} becomes

\begin{equation}				\label{Meqn}
M = m_{iso} - 25
	+ 2.5\log{f(z)}
	- 5\log{\left(\frac{D_L(z)}{\Mpc}\right)}
	- 2.5\log{k(z)}.
\end{equation}

\ni Including the fraction of detected light $f(z)$ has two effects.
First, because $f<1$, the absolute magnitude of each galaxy is
calculated to be brighter, which compensates for the unobserved light
beyond the limiting isophote.  Second, when equation \ref{Meqn} is
inverted to find the maximum redshift at which the galaxy would be
included in the sample (i.e.\ $z_{max}$ such that $m_{iso}=m_{lim}$),
then the effective luminosity distance is increased, and thus a
smaller $z_{max}$ is derived than in the absence of the lost light
term.  This reduces the maximum volume $V_{max}$, which in turn
increases the derived luminosity density (which scales with
$1/V_{max}$).

In luminosity function calculations which reduce the effects of
spatial inhomogeneities (Sandage et al.\ 1979, Efstathiou et al.\
1988), the derived luminosity function depends upon maximizing the
likelihood $\ln{\cal L}_i = \ln{\phi(M_i)} - \ln{
\int^{M_{max}(z_i)}_{M_{max}(z_i)} \phi(M) dM}$, summed over all
galaxies in the survey.  For these derivations, including the lost
light in equation \ref{Meqn} reduces the range of possible $M$, which
will change the shape of the luminosity function.  The normalization
of the luminosity function in these methods, like the classical
methods, depends upon the sum of $1/V_{max}$, and will also be higher
when the effects of lost light are included.

\section{Calculating the Observed Light Fraction $f(z)$}	\label{fsect}

Given equation \ref{Meqn} for the relationship between absolute
magnitude and apparent isophotal magnitude as a function of redshift,
it is in principle straightforward to calculate the luminosity function
for a given sample.  The reconstructions of the luminosity function
which are currently in use can readily be applied, but with the
substitution of the correct values of $z_{max}$, or with the correct
range of absolute magnitudes at a given redshift.

The difficulty in calculating the luminosity function becomes
calculating the form of $f(z)$ for the galaxies within a sample.  For
a galaxy with a fully specified two-dimensional surface brightness
profile, one can can calculate $f(z)$ at any redshift by first
rescaling the central surface brightness by $(1+z)^{-4}$ and applying
the $k$-correction, then calculating the angular size using the
angular diameter distance, and finally convolving the resulting 2-d
angular light distribution with the point spread function.  Then,
$f(z)$ may be calculated by integrating the resulting light
distribution over the area where the surface brightness is greater
than the limiting isophote $\mu_{lim}$, or within an area appropriate
to any other magnitude measure.

There are several existing papers which calculate the fraction of
detected light $f(z)$, particularly in the case of large photographic
surveys wherein image saturation is important but seeing can be
neglected.  Disney \& Phillipps (1983) have calculated $f(z)$ for
circular axisymmetric galaxies with both exponential and
de Vaucouleurs profiles, and used the resulting lost light fraction to
calculate the relative $V_{max}$ for galaxies of varying surface
brightness, at fixed total luminosity\footnote{The relative $V_{max}$
as a function of surface brightness, luminosity, and/or scale length
is sometimes referred to as the ``visibility'' in the literature
(e.g.\ Disney \& Phillipps 1983), since it reflects the relative
contribution of different galaxies to a given survey.  Thus, galaxies
which have a smaller $V_{max}$ are less likely to appear in a survey,
and are thus less ``visible''}.  Their analysis includes the effects
of combined isophotal magnitude limits and angular diameter limits.
Davies (1990) has expanded upon the work of Disney \& Phillipps, by
generalizing the calculation of the lost light and $V_{max}$ to
include combined bulge$+$disk models, and random inclinations.

The earlier work of Disney \& Phillipps has also been extended to
higher redshifts by Phillipps et al.\ (1990), who include the effect
of cosmological dimming and $k$-corrections in calculating both the
fraction of detected light and $V_{max}$ as a function of intrinsic
surface brightness and luminosity, for face-on exponential disks in
the absence of seeing.  Using a simple model, they demonstrate how
failure to compensate for the variation in the effective survey
volumes with variation in galaxy surface brightness can lead to
underestimating the luminosity function for low surface brightness
galaxies.

At moderate redshifts, however, seeing can be as important as surface
brightness in modifying the fraction of observed light and biasing
$V_{max}$.  When the half-light radius of a galaxy approaches the size
of the point spread function (PSF), the observed light profile becomes
dominated by the PSF.  This can becomes significant even at relatively
moderate redshifts ($z \sim 0.1$).  In a separate paper (Dalcanton
1998), we derive an analytic expression for the observed light
fraction $f(z)$ of axisymmetric exponential disks and de Vaucouleurs
profile ellipticals in the presence of a realistic Moffat (1969) point
spread function, which we will use for the duration of this paper;
Yoshii (1993) has calculated $f(z)$ for the simpler case of a Gaussian
point spread function.  We parameterize the two types of profiles with
a characteristic surface brightness $\mu_0$ and a length scale
$\alpha$.  For the exponential profile, $\mu_0$ is the central surface
brightness, and $\alpha$ is the exponential scale length.  For the de
Vaucouleurs profile, we chose $\alpha$ and $\mu_0$ such that the de
Vaucouleurs galaxy has the same half light radius and total flux as an
exponential galaxy with the same parameters.  With this choice,
$\mu_0=\mu_e-2.518$ and $\alpha=r_e/1.679$, where $\mu_e$ and $r_e$
are the more typical parameterizations of the de Vaucouleurs profile.

Using the prescription in Dalcanton (1998), and neglecting the
$k$-correction for generality, we have calculated the quantity $\Delta
M(z)=-2.5\log{f(z)}$, which characterizes the degree to which the
apparent isophotal magnitude $m$ falls off faster than $1/{D_L}^2(z)$.
We have plotted $\Delta M(z)$ for different galaxy sizes (solid,
dashed, and dotted lines) and characteristic surface brightnesses
relative to the limiting isophote ($\Delta\mu=\mu_{lim}-\mu_0$
decreasing from left to right), and for variable seeing (increasing
from top to bottom), for the cases of exponential (Fig.\
\ref{ffig}[a]) and de Vaucouleurs (Fig.\ \ref{ffig}[b]) light profiles.
These plots can be used to identify the regimes where a survey may
suffer from systematic biases in the measured the luminosity function;
when $\Delta M(z)$ is large over a survey's redshift range, the
apparent magnitude of a galaxy can be much fainter than would be
predicted by eqn.\ \ref{Morigeqn} alone, and thus, a given galaxy
drops below some isophotal magnitude limit faster than predicted by
eqn.\ \ref{Morigeqn}.

For both types of galaxy profiles, the curves
in Figure \ref{ffig} show the expected behavior for $\Delta M(z)$.  At
low redshifts, $\Delta M(z)$ is constant, because the fraction of lost
light does not change dramatically with increasing distance.  When the
redshift becomes large enough, however, a given galaxy becomes small
with respect to the point-spread function, and, if the redshift
approaches 1, the apparent surface brightness drops rapidly due to the
$(1+z)^{-4}$ surface brightness dimming.  These two effects drop the
apparent flux precipitously, increasing $\Delta M(z)$.

At low redshifts, where $\Delta M(z)$ is constant, the apparent
magnitude falls off as expected by eqn \ref{Morigeqn}, and thus, as
long as the maximum redshift reached by a survey falls within this
flat regime, the standard reconstruction of the luminosity function is
valid (i.e.\ eqn.\ \ref{Morigeqn})\footnote{For example, if one were
calculating the luminosity function for high surface brightness
galaxies selected from HST imaging (e.g.\ upper left in Figure
\ref{ffig}), the reconstructed luminosity function would be valid for
all redshifts less than 1.}.  In contrast, if the limiting magnitude
of a survey is very faint, then a significant fraction of galaxies
will be at large enough redshifts that $\Delta M(z)$ is not constant
over the survey volume.  In this case, the apparent magnitude will be
falling faster than expected, and thus the luminosity function will be
systematically underestimated.  In particular and perhaps
counterintuitively, the luminosity function is more likely to be
underestimated in the bright end, since these galaxies will be
detectable over the largest range in redshift.  On the other hand, the
redshift at which $\Delta M(z)$ becomes large is smallest for
intrinsically small galaxies, and for low surface brightnesses
galaxies.  Thus, in deep enough surveys, the luminosity function may
be biased in the faint end.  The error in the derived luminosity
function will also depend upon the characteristic redshift of the
survey, and thus, the luminosity function calculated for the same type
of galaxies can be quite different from survey to survey, due to
variations in the maximum redshift explored by a survey (i.e.\
$m_{lim}$), variations in the seeing, or variations in the limiting
isophote $\mu_{lim}$.

Even in the case of optimal seeing, where $\Delta M(z)$ is constant
over a large redshift range, there can be large effects on the derived
luminosity function.  As seen if Figure \ref{ffig}, when a galaxy's
central surface brightness approaches the limiting isophote, there is
a global offset in $\Delta M(z)$, which increases as the surface
brightness decreases.  This offset leads low surface brightness
galaxies to have systematically underestimated luminosities (see
discussions by Disney 1976, Disney \& Phillipps 1983, for example),
even in the absence of seeing effects.  Thus, only the very nearest
low surface brightness galaxies will be bright enough to be within the
magnitude limit of a survey, and as such, LSBs have much lower
detection volumes, even at low redshifts where $\Delta M(z)$ is
constant; this argument has been framed as the classic ``visibility''
problem by Disney (1976), Disney \& Phillipps (1983), Davies (1990),
\& McGaugh et al.\ (1995).  Furthermore, if surface brightness is
correlated with absolute magnitude, as seen in clusters of galaxies
(e.g.\ Binggeli et al.\ 1984), then the underestimate of the
luminosity becomes increasingly severe at faint absolute magnitudes.
This correlation would produce a stretching of the luminosity
function, which in turn would lead to a smaller faint end slope.  The
effect is nearly independent of seeing, and has been discussed by
Phillips et al.\ (1990).

Figure \ref{ffig} also reinforces the value of having very
deep, high-resolution imaging data when selecting targets for a
redshift survey, such that the magnitude limit of the survey
is many magnitudes above the limiting magnitude of the imaging
data.  For such cases, the vast majority of galaxies will
have limiting isophotes for both detection and photometry which
are well below their central surface brightness, and as Figure \ref{ffig}
shows, such surveys will have far fewer problems with isophotal
magnitude biases, as they will contain a very large fraction of
most galaxies' light out to $z\sim1$.

Finally, to give some appreciation for the values of
$\Delta\mu=\mu_{lim}-\mu_0$ and $\alpha$ which might be applicable for
typical surveys, Figure \ref{mualphafig} shows $\mu_0$ and $\alpha$
derived from surface photometry for an assortment of galaxy samples,
chosen to be representative of the range of galaxy morphologies and
luminosities (and not to be a definitive compilation on galaxy
structural parameters).  Bright ellipticals tend to have $21<\mu_0<
18\,B\surfb$, while spirals have $\mu_0>20\,B\surfb$, down to the
current surface brightness limits of major surveys
($\mu_0\sim24\,B\surfb$).  Bulges tend to have the highest surface
brightnesses ($15<\mu_0<21\,B\surfb$), but also tend to be factors of
10 smaller than ellipticals and spirals.  However, the lack of small
spirals with $\alpha<1\,h_{50}^{-1}\kpc$ is an artifact of only
plotting spirals drawn from angular diameter limited surveys, which
are strongly biased towards identifying galaxies with large physical
sizes.  We also note that, because of $k$-corrections and evolutionary
effects, the appropriate value of $\mu_0$ can be a strong function of
redshift; for example, while elliptical galaxies might have
$\Delta\mu=5$ in blue bandpasses at low redshift, a value of
$\Delta\mu=1$ may be more appropriate at redshifts close to 1.  There
are fewer surface photometry measurements available for other band
passes, but for reference, the equivalent to the Freeman (1970)
surface brightness is $20.1\surfb$ in $r$, and $\sim17.6\surfb$ in $K$
for galactic disks (Courteau 1996, de Jong 1996b).

To derive $\Delta\mu$, the surface photometry presented in Figure
\ref{mualphafig} can be compared with the isophotal limits for recent
field surveys, which we have attempted to summarize in Table
\ref{surveytab}.  The Table is provided only as a rough guide, given
that in many cases, the photometric selection criteria are complicated
and$/$or not fully specified, and the reader should refer to the
original papers for a more complete description.  For many of the
entries we have estimated the limiting surface brightness for
detection based upon the given photometric limits, or measures of the
sky noise, where possible; only the CFRS and Norris surveys have
adequately tested this limit using realistic artificial galaxy tests.
An examination of Table \ref{surveytab} shows that the best infrared
field surveys to date have only $\Delta\mu\lta1.5$ for pure spiral
disks, suggesting that the current generation of $K$ band luminosity
functions reflect the properties of the high surface brightness
elliptical population alone.

\section{Luminosity Functions}			\label{lumfuncsect}

The above results suggest that the measured luminosity function can
depend sensitively upon the survey used to select galaxies, and upon
the redshifts and morphologies of the galaxies which have been
selected.  In this section we explore in more detail the specific ways
in which the luminosity function may be affected, by simulating the
entire process of measuring a luminosity function, including both
sample selection and standard reconstructions of the luminosity
function.  For this exercise, we will neglect spatial density
inhomogeneities, and use the ``classical'' $1/V_{max}$ calculation for
deriving the luminosity function; more sophisticated
maximum-likelihood methods should produce identical results in the
limit of smooth galaxy distributions.

In general, reconstruction of the luminosity function proceeds as follows:

\ni Step 1. A sample of galaxies whose apparent magnitude $m$ is above
some threshold $m_{lim}$ is selected.  There may also be additional
selections for apparent angular size, particularly for nearby
photographic surveys.  Size criteria will be neglected for the
following exercise.

\ni Step 2. Redshifts are measured for the selected galaxies, and,
when combined with the apparent magnitudes $m$, are used to derive
absolute magnitudes using equation \ref{Morigeqn}.

\ni Step 3. The maximum redshift possible for each galaxy is calculated by
solving equation \ref{Morigeqn} for $z_{max}$, using the derived
absolute magnitude from Step 2, and assuming $m=m_{lim}$.

\ni Step 4. The volume $V$ within $z_{max}$ is calculated for each galaxy,
and then the inverse is summed over all galaxies within absolute
magnitude bins, to derive the relative number density as a function of
absolute magnitude.

There are two competing effects which can affect the galaxy luminosity
function derived with the standard methods.  First, due to the
lost-light beyond the outer isophote, galaxies will have
systematically fainter derived absolute magnitudes, because Step 2
assumes that the observed magnitude $m$ is a good representation of
the total flux from a galaxy.  Then, when the artificially faint
absolute magnitude is used to calculate $z_{max}$ in Step 3, one will
estimate that the galaxy has a smaller maximum redshift than if all
the light were included in the measurement of $m$.  This leads one to
underpredict the maximum volume $V_{max}$, and thus calculate larger
values of the luminosity function in Step 4.  One will also calculate
larger values of $V/V_{max}$ for galaxies which are strongly affected
by this bias.  The second effect works in the opposite sense.  As
shown in \S\ref{fsect}, due to the variation in the fraction of
detected light $f(z)$ with redshift, a galaxy's apparent magnitude can
fall off much faster with redshift than predicted by equation
\ref{Morigeqn} alone (i.e.\ eqn.\ \ref{Meqn}).  This leads one to
overpredict the maximum redshift, and thus underpredict the luminosity
function.  In this case, the derived values of $V/V_{max}$ will be
smaller than expected.  Because these two effects compete with each
other, it is possible for a sample to have $<V/V_{max}>\sim0.5$, in
spite of the measured luminosity function being strongly distorted
from the true underlying distribution.

To demonstrate the size of these effects, we have modelled the case of
a moderately deep redshift survey in Figure \ref{lffig}, for the
specific case of a limiting isophotal magnitude of $m_{lim}=22$
within a limiting isophote of $\mu_{lim}=25\surfb$, for galaxies with
different characteristic surface brightnesses, and for both
exponential and de Vaucouleurs light profiles.  The analysis assumes
that one is calculating the full bivariate luminosity
function\footnote{This has also been referred to at the bivariate
brightness function (BBF) by Phillipps et al.\ (1990) and Boyce \&
Phillipps (1995).  We prefer to retain the phrase ``luminosity
function'', to maintain the analogy to previous work on the galaxy
luminosity function.  See Boyce \& Phillipps (1995) for a good
discussion of the elements necessary to a a survey designed to measure
the full bivariate luminosity function}, where the galaxies are binned
not only in luminosity, but also in intrinsic surface brightness
$\mu_0$.

To create a simulated galaxy sample, we have populated a universe with
a uniform density of face-on galaxies of a fixed characteristic
surface brightness, drawn randomly from a Schechter (1976) luminosity
function $\Phi(M)$ with $M_*=-21$ and $\alpha_{lum}=-1.5$.  Note that
the choice of $M$ and $\mu_0$ fixes the exponential scale length, such
that variation in $M$ at fixed surface brightness is effectively a
variation in size.  This is a reasonable approach for spiral disks,
which span a continuous range of $\mu_0$ and $\alpha$, but is somewhat
artificial for pure elliptical galaxies, which tend to fall along a
locus in $\mu_0$ vs $\alpha$ (Figure \ref{mualphafig}).  We retain
this choice for ellipticals, however, because it more clearly
elucidates the general trends which are expected, and because, in
general, a bivariate luminosity function will probably not further
subdivide a luminosity function by profile shape, in addition to
surface brightness.  For each surface brightness, we calculate the
observed isophotal magnitude using Moffat convolved galaxy profiles
and generate a catalog of 5,000 galaxies which fall above the
isophotal magnitude limit.  The resulting catalog is then analyzed
using the methods most typically applied to galaxy catalogs (Steps 1-4
above; eqn.\ \ref{Morigeqn}), and then reanalyzed using the proper
correction for lost light (eqn.\ \ref{Meqn}).  For the sake of
generality, we have neglected the $k$-correction term in eqn.\
\ref{Meqn}, both in generating the catalog and in analyzing it; we
will discuss the role of the $k$-correction below.  The final
luminosity function and $V/V_{max}$ distributions are shown in Figure
\ref{lffig}.


There are several obvious trends demonstrated in the luminosity
functions plotted in Figure \ref{lffig}:

1. As expected from Figure \ref{ffig}, the errors in the derived
luminosity function are greater for low surface brightness galaxies,
due to the increased fraction of lost light, and the lower redshift at
which $\Delta M(z)$ becomes large.  The biases are slight
when the characteristic surface brightness $\mu_0$ is $5\surfb$ above
the limiting magnitude, significant at $\Delta\mu=3\surfb$, and severe
at $\Delta\mu=1\surfb$ (where $\Delta\mu=\mu_{lim}-\mu_0$).

2. The underestimate in the luminosity function is worse at high
redshift, due to the rapid variation in the lost light fraction at
increasing redshift.  Furthermore, in low-redshift subsamples, some
portion of the sample has $V_{max}$ fixed by the upper cutoff in
redshift, which reduces the impact of ignoring the lost light
corrections.  This suggests that when the luminosity function is
derived in intervals of redshift, evolution detected in the luminosity
function may be correct for the lower redshift intervals, but
evolution between the penultimate and the highest redshift interval is
suspect.

3. The underestimate of the luminosity function is worse at bright
magnitudes, as these galaxies tend to be at larger redshifts, where
the light lost fraction changes rapidly.

4. The luminosity functions appear to shift to fainter $M_*$ with
decreasing surface brightness, due to the increasing offset in $\Delta
M(z)$ (see Figure \ref{ffig}).

5. The mismeasurement of the luminosity function can mimic luminosity
and density evolution.  The luminosity function appears to shift
towards brighter magnitudes and to larger normalizations as the
redshift decreases (see the comparison between the dotted and solid
curves in the lower left panels of Figure \ref{lffig}).  Including
$k$-corrections and evolutionary corrections can induce the
opposite effect.  Most galaxies are probably intrinsically brighter at
higher redshifts, and thus the appropriate value of $\Delta\mu_0$ may
be larger with increasing redshift, if the evolutionary corrections are
larger than the $k$-corrections (as might be the case for galaxies
with an exponentially declining star-formation rate).  The luminosity
function will then shift to higher luminosities at larger redshifts
due solely to biases, and will therefore have the appearance of more
rapid luminosity evolution than is the case.

6. For samples with the same number of galaxies, the underlying
luminosity function has a much higher normalization for low surface
brightness galaxies, due to their smaller mean distance in isophotal
magnitude limited surveys; note the difference in the mean redshift as
the surface brightness is decreased.  This is the classical
``visibility'' effect discussed by Disney (1976), Disney \& Phillips
(1983), Davies (1980), McGaugh et al.\ (1995), and others.  The
relative normalizations of the luminosity functions suggest that
galaxies with $\Delta\mu=1\surfb$ will be underrepresented by a factor
of 10, compared to a population of galaxies with an identical space
density and $\Delta\mu=5\surfb$.  If the lower surface brightness
galaxies are not treated separately, then the derived luminosity
function will be 10\% higher than would be correct for
$\Delta\mu=5\surfb$, but will be a factor of 10 smaller than is
correct for the entire surface brightness range spanned by the sample
($\Delta\mu>1\surfb$).  In such instances, one should quote that the
derived luminosity function is representative of only the high surface
brightness fraction of the sample, even if the survey is capable of
reliably detecting galaxies at much lower surface brightnesses.

7. At the same surface brightness, the mismeasurement of the
luminosity function is comparable for both exponential and
de Vaucouleurs profiles, as would be expected by the general similarity
of the detected light fraction $f(z)$ plotted in Figure \ref{ffig}.
However, ellipticals and bulges have somewhat higher
surface brightnesses than spiral disks (Figure \ref{mualphafig}), by
about 1-2$\surfb$ at the bright end, and thus ellipticals and early
type spirals will be tend to be less affected by luminosity function
biases, when the $k$-correction is negligible.  In contrast, Figure
\ref{mualphafig} also shows that ellipticals and spirals have
comparable half-light radii, and thus both types of galaxies will
begin to be affected by seeing at similar redshifts.  Bulges have much
smaller half-light radii than their spiral hosts, and while a high
surface brightness bulge will help a spiral galaxy's detectability, at
redshifts $z\gta0.1$ the bulge profiles begin to be strongly affected
by seeing for typical ground based observations.  As a result, bulges
may tend to become undetectable at comparable redshifts as their
associated disks (e.g.\ compare the tracks in Figure \ref{ffig}[b] for
$\Delta\mu=7$ and $\alpha=0.25\kpc$ to the tracks in Figure
\ref{ffig}[a] for $\Delta\mu=3$ and $\alpha=3\kpc$).

8. There is less appearance of ``evolution'' for the de Vaucouleurs
profiles than for the exponentials, in the absence of the
$k$-correction, because $f(z)$ is constant over a slightly larger
range in redshift for the de Vaucouleurs profile, while having a sharper
increase at high redshift, and because $f(z)$ varies more strongly
with surface brightness in exponential disks than in de Vaucouleurs
ellipticals.  These effects also lead to ellipticals having smaller
mean redshifts than comparable spirals at high surface brightnesses
($\Delta\mu>3$) and larger mean redshifts at low surface brightnesses
($\Delta\mu\le1$).

\section{$V/V_{max}$ Distributions}			\label{Vsect}

The degree to which the luminosity functions seen in Figure
\ref{lffig} have systematic problems can be partially diagnosed using
the derived $V/V_{max}$ distributions.  As presented in Figure
\ref{lffig}, as the corrections for lost light become larger (i.e.\ at
lower central surface brightnesses and higher redshifts) the
distributions of $V/V_{max}$ become sloped, such that the distributions
are elevated at small values of $V/V_{max}$ and are depressed at larger
values.  The nature of the deformation is such that it nearly preserves the
expectation value of $<V/V_{max}>=0.5$, even in the presence of severe
problems in the derivation of the luminosity function.  Thus, the full
distribution of $V/V_{max}$ is needed to test for the presence
of lost-light biases.  In the presence of the necessary $k$-correction,
however, the full $V/V_{max}$ distribution may well deviate from the
relatively straight line seen in Figure \ref{lffig}, and pull
the mean $<V/V_{max}>$ from its expected value of 0.5.

Ideally, the $V/V_{max}$ test should be performed in several
redshift bins, given that the low redshift subsample of the
$\Delta\mu=1\surfb$ sample is clearly biased in its luminosity
function derivation, while having a nearly normal $V/V_{max}$
distribution.  This apparent normality results from $V_{max}$ being
fixed by the upper cutoff in redshift for a large fraction of the
galaxies.  In general, the slope in the $V/V_{max}$ distribution is
somewhat more severe for disk galaxies than comparable elliptical
galaxies.

We note that the expected signature has already been seen in the data
used for the Autofib survey (Ellis et al.\ 1996; their Figure 2).
Even after correction for redshift incompleteness, their published
$V/V_{max}$ distributions systematically fall off towards large values
of $V/V_{max}$, and look remarkably similar to the plots in Figure
\ref{lffig}.  Most other existing surveys have published only the mean
$<V/V_{max}>$ and cannot be immediately tested for the biases
discussed here. 

\section{Alternative Measures of Luminosity}		\label{petrosiansect}

The systematic biases in the luminosity functions presented
in Figure \ref{lffig} are a direct result of the practice
of using isophotal magnitudes to measure galaxies' luminosities.
There are other ways to measure galaxy magnitudes however, which,
while less commonly used, can potentially improve the accuracy of
the luminosity function reconstruction.

The easiest alternative to the isophotal magnitude is the aperture
magnitude.  One can choose a large enough aperture to safely contain a
large fraction of a typical galaxy's light, and thus have a measure of
the total luminosity which is not biased with surface brightness.
However, aperture magnitudes are strongly biased with galaxy size;
physically large galaxies will have a smaller fraction of their light
detected than will smaller galaxies, leading the brighter galaxies to
be measured with a smaller intrinsic luminosity.  The degree of
mismeasurement will also be a function of redshift, with a larger
fraction of a galaxies' light being detected at higher redshift.  This
redshift dependence is opposite to the behavior of isophotal
magnitudes, as can be seen in the top row of Figure
\ref{petrosianfig}, where we have plotted the error in the measured
absolute magnitude as a function of redshift, for a $3\arcsec$
diameter aperture (as in Lilly et al.\ 1991, for $1\arcsec$ seeing),
applied to various galaxy profiles and physical sizes.  Because 
it is the redshift dependence of $\Delta M$ which led to the artificial
appearance of galaxy evolution in Figure \ref{lffig}, aperture
magnitudes should also produce artificial evolution, although with
the opposite luminosity evolution.  As can be seen from Figure
\ref{petrosianfig}, even for a large aperture ($3\times$FWHM),
a typical spiral with $\alpha=3\kpc$ is biased by over 1 magnitude
a $z=0.1$.

Another common practice is to attempt to correct a measured isophotal
magnitude to a ``total'' magnitude.  The ``total'' magnitudes provided
by the FOCAS package (Valdes 1982) are created by extending the
isophotal area by a factor of 2, and then measuring the flux within
the larger aperture.  While not immediately apparent, the FOCAS
``total'' magnitudes are actually equivalent to an isophotal magnitude
measured at a fainter limiting isophote.  Thus, one can consider using
a FOCAS ``total'' magnitude as increasing the value of
$\Delta\mu=\mu_{lim}-\mu_0$, where $\mu_{lim}$ is the limiting
isophote for the initial detection of the galaxy.  For pure
exponentials, the new effective magnitude difference is always
$\Delta\mu_{eff}=\sqrt{A}\Delta\mu$, where $A$ is the factor by which
the isophotal area is grown.  For pure de Vaucouleurs profiles,
$\Delta\mu_{eff}=A^{1/8}\Delta\mu+5.81(A^{1/8}-1)$, and for galaxies
which are small enough to have nearly Gaussian profiles due to the
seeing, $\Delta\mu_{eff}=A\Delta\mu-5(A-1)\log{\sigma/\alpha}$, where
$\sigma$ is the standard deviation of the Gaussian point spread
function, and $\alpha$ is the galaxy's scale length.  Because the
increase in $\Delta\mu_{eff}$ is largely a multiple of the isophotal
value of $\Delta\mu$, changing from an isophotal magnitude to a
``total'' magnitude does not greatly reduce the level of systematic
bias for lower surface brightness galaxies; a galaxy with
$\Delta\mu=1.5$ has $\Delta\mu_{eff}=1.5-2.1$ when FOCAS ``total''
magnitudes are used.

There are other forms of ``total'' magnitudes, besides those used by
FOCAS.  One of the most common of these is the Kron (1980) magnitude,
which sets the aperture to be a multiple of the intensity-weighted
first moment radius $r_1$, instead of a multiple of the isophotal
area.  With careful consideration of the effects of seeing, the Kron
magnitude can provide a fairly robust method of providing a magnitude
measurement which is consistent over a range in redshifts and
observing conditions (see Bershady et al.\ 1994 for a particularly
cogent discussion of the relevant issues).  The consistency of the
Kron magnitude with variations in surface brightness has not yet been
sufficiently explored, however.  Galaxies whose apparent central
surface brightness approaches the isophotal limit are typically
detected out to only one scale length, and thus will tend to give
systematically smaller values of $r_1$; the degree of this
mismeasurement will depend sensitively on the details of how $r_1$ is
measured (iteratively, directly, etc).

We test the sensitivity of Kron magnitudes to variations in surface
brightness in Figure \ref{kronfig}, where we have assumed that $r_1$
is measured above some limiting isophote $\mu_{lim}$, and that the
Kron magnitude is calculated within a radius $r_{ap}=2r_1$, as
suggested in Kron (1980).  Figure \ref{kronfig} shows that the Kron
magnitude underestimates the total luminosity at low surface
brightnesses ($\mu_{lim}-\mu_0\lta3$), much like isophotal magnitudes
do.  For $\Delta\mu\sim3$, errors in the absolute magnitude are larger
than $0.3-0.7^m$, and increase in size towards fainter surface
brightnesses (i.e.\ smaller $\Delta\mu$).  The errors at low surface
brightnesses are also larger when the PSF dominates the galaxy
profile.  The curves shown in Figure \ref{kronfig} are actually worst
case scenarios, as we do not calculate $r_1$ iteratively, nor do we
prevent the Kron radius $2r_1$ from becoming smaller than either some
multiple of the original limiting isophotal radius or some fixed
aperture radius, both of which will give better estimates of the total
magnitude than given in Figure \ref{kronfig}.  However, as in the case
of FOCAS total magnitudes, calculating $r_1$ iteratively is equivalent
to using a fainter value of $\mu_{lim}$, and thus a somewhat larger
value of $\Delta\mu$.  The curves in Figure \ref{kronfig} will
therefore be identical when iterative measures of $r_1$ are used, but
the appropriate value of $\Delta\mu$ will be larger.  The situation is
similar when there is a minimum allowed Kron radius, if the minimum
radius is set to be a multiple of the original limiting isophotal
radius (as in SExtractor, Bertin \& Arnouts 1996).  If the minimum
Kron radius is a fixed aperture, then the biases for the faintest
galaxies will become those suffered by aperture magnitudes.

Another alternative to either the isophotal or aperture magnitude is
the Petrosian magnitude (1976).  To define a Petrosian magnitude, one
fixes the value of the parameter $\eta$, finds the Petrosian radius
where $\eta = <\Sigma(r)>/\Sigma(r)$, and then measures the flux
within that radius (here, $<\Sigma(r)>$ is the mean surface brightness
within $r$).  Because the characteristic surface brightness cancels in
the definition of $\eta$, the Petrosian radius will be identical for
galaxies with the same profile shape, independent of their surface
brightness, and thus the fraction of detected light $f(r)$ will be
constant with surface brightness as well.  Furthermore, unlike for
aperture magnitudes, the fraction of a galaxy's luminosity contained
within the Petrosian radius is independent of a galaxy's size or
distance, and depends only upon the particular choice of $\eta$, and
the shape of the galaxy's profile (in the absence of seeing).  These
qualities make the Petrosian magnitude a nearly ideal choice for
measuring luminosity functions in a way that is unbiased with redshift
or surface brightness.  The stability of the Petrosian magnitude can
be seen in the bottom of Figure \ref{petrosianfig}.  The Petrosian
absolute magnitude is nearly constant with redshift for pure
exponential disks of different sizes, even when a $1\arcsec$ PSF is
included.  For de Vaucouleurs profiles, the absolute magnitude varies by
only 0.2 magnitudes out to a redshift of 2, and differs from
exponential profiles by roughly the same amount.  We have chosen the
value of $\eta$ carefully, to give a good compromise between including
a large fraction of a galaxy's light, yielding consistent results for
elliptical and spiral profiles, and producing small enough Petrosian
radii that $\Sigma(r)$ can be accurately measured; for different
choices of $\eta$, the results in Figure \ref{petrosianfig} could be
very different.  The comparison of exponential profiles to
de Vaucouleurs profiles in Figure \ref{petrosianfig} suggests that even
for a realistic mixture of galaxies, Petrosian magnitudes should be
able to make consistent measures of luminosity over a large range in
galaxy size, surface brightness, and redshift, leading to a vast
improvement over the standard use of isophotal magnitudes.

While Petrosian magnitudes appear to offer a nearly ideal tool for
measuring galaxy luminosities, the plots in Figure \ref{petrosianfig}
disguise several difficulties.  In general, Petrosian magnitudes are
subject to large errors at low signal-to-noise ratios and for poorly
sampled profiles.  To find the radius corresponding to a particular
value of $\eta$ in a faint galaxy image, one must carefully
interpolate over a relatively small number of pixels to find $r$,
$\Sigma(r)$ and $<\Sigma(r)>$.  For low surface brightness galaxies or
normal galaxies at moderate redshifts, the particular value of
$\eta$ may require accurately measuring extremely faint surface
brightnesses (i.e.\ $\mu_0 \sim 30\surfb$), which may
be well beyond the capabilities of the available imaging data, due to
sky noise, flat-fielding, crowding, etc.  Thus, while Figure
\ref{petrosianfig} suggests that Petrosian magnitudes are unbiased
with surface brightness or size in ideal data, for noisy, binned data,
there should be both noise and systematic biases which will get worse
with decreasing surface brightness and redshift, although hopefully to
a lesser degree than isophotal magnitudes.

Wirth (1996) has explored the systematic errors in Petrosian
magnitudes for different interpolation schemes, and for different
signal-to-noise ratios.  He finds that the simplest interpolation
methods are highly biased with signal-to-noise, and while more
sophisticated ones can be constructed, they are only useful for
galaxies which have a signal-to-noise ratio above 10 within their true
half-light radii.  At SNR$_{0.5}=10$, the best estimators have a
typical scatter in the Petrosian radius of more than 50\%, and a
systematic bias of about 30\%, which will translate into
systematic errors and increased scatter in the measured Petrosian
magnitudes for faint galaxies.  The simplest interpolation schemes
produce much larger systematic biases, which reach over 80\% in
the Petrosian radius at SNR$_{0.5}=10$.  Because of these systematics,
the robustness of any adopted scheme for measuring Petrosian
magnitudes should be thoroughly tested as a function of surface
brightness, size, and signal-to-noise, before they can be safely
invoked as a panacea for the problems which can plague isophotal
magnitude determinations of the luminosity function.

Wirth (1996) also finds that for galaxies with high signal-to-noise
(SNR$_{0.5} \sim 80$), the measured Petrosian radii are good to within
10\% which suggests that one could perform a nearly ideal luminosity
function reconstruction for deep enough imaging data.  However, even
the best data is subject to the strong isophotal magnitude biases
discussed in the rest of this paper.  The detection of galaxies is
fundamentally an isophotal signal-to-noise limited process; one will
never measure photometry for a galaxy which does not have substantial
light above the sky noise of a given image.  Thus, all surveys have an
implicit isophotal magnitude limit, which is roughly the standard
completeness limit of a survey.  In determining the biases in a
luminosity function measurement, even if using Petrosian magnitudes,
one must also consider the imact of the implicit isophotal magnitude
limit as well.

\section{Conclusions and Complications}			\label{conclusions}

In the current epoch of deep CCD imaging, some surveys have
extended into a redshift regime where the light lost beyond the outer
detection isophote, due to intrinsic surface brightness and seeing,
can have large, systematic effects upon the derived luminosity
function, without violating the standard $<V/V_{max}>$ tests which are
often used to diagnose incompleteness and unusual systematics.  We
have attempted to give a general guide to determining the regimes
where corrections for lost light become important, and to estimating
the degree to which neglecting the corrections can affect the
resulting luminosity function.  We have also shown how these effects can
easily mimic evolution in the galaxy population.  As an aside, we
note that the biases we discuss are potentially relevant to any
work which involves detection of extended objects, such as the
detection of galaxy clusters in X-ray data.

In our investigation, we have taken the simplest, most general
assumptions, in order to clarify the general biases which affect
luminosity function measurements, and to estimate the size of these
biases in different regimes.  This approach comes at the expense of
glossing over some subtleties which can affect real surveys.  Perhaps
the most glaring complication is that in many surveys, the limiting
isophote for detection is substantially brighter than the limiting
isophote for photometry.  For example, the CFRS survey (Lilly et al.\
1995a) uses a very deep isophotal limit of roughly $\mu_{I_{AB}}=28.5$
for photometry (which guarantees that they measure a consistently
large fraction of a detected galaxy's light), but has an effective
isophotal limit for detection of roughly $\mu_{I_{AB}}\sim25.5$
(estimated from the central surface brightness detection limit in
their Figure 2), due to sky noise and flat-fielding uncertainties.  In
such cases, one must take additional care in calculating $V_{max}$,
such that the maximum redshift is derived with the additional criteria
that the object can be detected above the limiting isophote for
detection, as well as having an isophotal magnitude brighter than the
limiting magnitude of the survey.  Testing for a galaxy's
detectability is further complicated by the usual practice of
smoothing an image before searching for galaxies within it.  For
imaging data which is substantially deeper than the limiting magnitude
of the spectroscopic subsample, many of these problems can be avoided.

For simplicity, we have also neglected the mix of complicated
morphologies and viewing angles present in the real universe, and
instead used face-on circularly symmetric exponential disks and
de Vaucouleurs profile ellipticals for illustration.  As almost all
galaxies can be reasonably approximated by a sum of the two profile
types, reality is likely to be bracketed by these two limiting cases.
If a survey is in a redshift regime or a surface brightness
regime where the corrections for lost light are probably large (see
Figure \ref{ffig}), then the optimal method for
deriving $V_{max}$ is to first fit an inclined bulge/disk model to
each galaxy, then to calculate $V_{max}$ by simulating the individual
galaxies as they would be observed at higher redshifts (including the
likely $k$-corrections), using either analytic calculations or
artificial galaxy tests.  In the latter case, one can readily
incorporate the complications of galaxy detection as a function of
redshift, as well as the effects upon the photometry.  We have
also shown how the use of Petrosian magnitudes can greatly reduce
the size of the needed corrections, but have argued that their benefits
are most substantial at relatively high signal-to-noise ratios.

While making these further corrections can be a fair bit of work, and
will certainly be subject to their own errors and biases, for the
typical size of galaxy surveys ($\lta1000$ galaxies) such an approach
is not necessarily untenable, and for larger galaxy surveys, such as
the upcoming 2DF and Sloan surveys, this can be approached in a
binned, statistical way, and/or can be heavily automated.  Many groups
have already implemented routines to estimate structural parameters
for galaxies using deconvolution and/or maximum likelihood methods
(e.g. Schade et al.\ 1996, Ratnatunga et al.\ 1994), and from there,
it is a small extrapolation to calculate the correct $V_{max}$.  At
the least, it is preferable to make a first-order correction when
needed, than to ignore effects which may be large enough to obscure
the cosmology that one is hoping to illuminate.

\bigskip
\centerline{Acknowledgements}
\medskip

It is a pleasure to thank Rebecca Bernstein for sage council during
the course of this work, Steve Shectman for his services as a sounding
board in the early stages, Dan Rosenthal for shrewd discussions, and
John Mulchaey for the tireless efforts of his workstation.  An
anonymous referee is also thanked for many helpful suggestions.
Support was provided by NASA through Hubble Fellowship grant \#2-6649
awarded by the Space Telescope Science Institute, which is operated by
the Association of Universities for Research in Astronomy, Inc., for
NASA under contract NAS 5-26555.

\section{References}

\hi{Arp, H. 1965, \apj 142, 402.}

\hi{Bershady, M. A., Hereld, M., Kron, R. G., Koo, D. C., Munn, J. A., \&
Majewski, S. R. 1994, \aj 108, 870.}

\hi{Bertin, E., \& Arnouts, S. 1996, \aasup 117, 393.}

\hi{Binggeli, B., Sandage, A., \& Tammann, G. A. 1988, \araa 26, 509.}

\hi{Binggeli, B., Sandage, A., \& Tammann, G. A. 1984, \aj 89, 64.}

\hi{Boroson, T. 1981, \apjs 46, 177.}

\hi{Boyce, P. J., \& Phillipps, S. 1995, \aa 296, 26.}

\hi{Caldwell, N., Armandroff, T. E., Seitzer, P., \& Da Costa, G. S., 1992
\aj 103, 840.}

\hi{Courteau, S., 1996, \apjs 103, 363.}

\hi{Dalcanton, J. J. 1998, \aj submitted.}

\hi{Davies, J. I. 1990, \mn 244, 8.}

\hi{de Jong, R. S., 1996a, \aasup, 118, 557.}

\hi{de Jong, R. S., 1996b, \aa, 313, 45.}

\hi{Disney, M. 1976, \nat 263, 573.}

\hi{Disney, M., \& Phillipps, S. 1983, \mn 205, 1253.}

\hi{Efstathiou, G., Ellis, R. S., \& Peterson, B. A. 1988, \mn 232, 431.}

\hi{Ellis, R. S., Colless, M., Broadhurst, T., Heyl, J., \& Glazebrook,
K. 1996, \mn 280, 235.}

\hi{Ferguson, H. C., \& McGaugh S. S. 1995, \apj 440, 470.}

\hi{Freeman, K. 1970, \apj 160, 811.}

\hi{Glazebrook, K., Peakcock, J. A., Colins, C. A., Miller, L. 1994,
\mn 266, 65.}

\hi{Gardner, J. P., Sharples, R. M., Carrasco, B. E., \& Frenk, C. S. 1996,
\mn 282, L1.}

\hi{Humason, M. L., Hayall, N. U., \& Sandage, A. R. 1956, \aj 61, 97.}

\hi{Jorgensen, I., Franx, M., \& Kjaergaard, P. 1995, \mn 273, 1097.}

\hi{Knezek, P. 1993, Ph.D. Thesis, University of Massachusetts.}

\hi{Kron, R. G. 1980, \apjs 43, 305.}

\hi{Lilly, S. J., Cowie, L. L., \& Gardner J. P. 1991, \apj {\bf369}, 79.}

\hi{Lilly, S. J., Le F\'evre, O., Crampton, D., Hammer, F., \& Tresse, L.
1995a \apj 455, 50.}

\hi{Lilly, S. J., Tresse, L., Hammer, F., Crampton, D., Le F\'evre, O.
1995b \apj 455, 108.}

\hi{Lin, H., Kirshner, R. P., Shectman, S. A., Landy, S. D., Oemler, A.,
Tucker, D. L., \& Schechter P. L. 1996, \apj 464, 60.}

\hi{Lin, H., Yee, H. K. C., Carlberg, R. G., \& Ellingson, E.
1997, \apj 475, 494.}

\hi{McGaugh, S. S. 1994, \nat 367, 538.}

\hi{McGaugh, S. S., \& Bothun, G. D. 1994, \aj 107, 530.}

\hi{McGaugh, S. S., Bothun, G., \& Schombert, J. 1995, \aj 110, 573.}

\hi{Moffat, A. F. J. 1969, \aa 3, 455.}

\hi{Petrosian, V. 1976, \apjl, 209, 1.}

\hi{Phillipps, S., Davies, J. I., \& Disney, M. J. 1990, \mn 242, 235.}

\hi{Phillipps, S., \& Driver, S. 1995, \mn 274, 832.}

\hi{Ratnatunga, K. U., Griffiths, R. W., \& Casertano, S. 1994, in
The Restoration of HST Images and Spectra--II,
ed. R. J. Hanisch \& R. L. White (Baltimore: STScI), 222.}

\hi{Reeves, G. 1956, \aj 61, 69.}

\hi{Romanishin, W., Strom, K. M., \& Strom, S. E. 1983, \apjs 53, 105.}

\hi{Sandage, A., 1961, \apj 133, 355.}

\hi{Sandage, A., Tammann, G. A., \& Yahil, A. 1979, \apj 232, 352.}

\hi{Schechter, P. 1976 \apj 203, 297.}

\hi{Schade, D., Lilly, S. J., Le Fevre, O., Hammer, F., \& Crampton,
D. 1996, \apj 464, 79.}

\hi{Schombert, J. M., Bothun, G. D., Schneider, S. E., \& McGaugh, S. S. 1992,
\aj 103, 1107}

\hi{Shectman, S., Landy. D., Oemler, A., Tucker, D., Lin, H., Kirshner, R., \&
Schechter, P. 1996, \apj 470, 172}

\hi{Small, T. A., \& Sargent, W. L. W. 1997, \apjs, 111.}

\hi{Valdes, F. 1982, {\it S.P.I.E.}, 331, 465.}

\hi{de Vaucouleurs, G., de Vaucouleurs, A., Corwin, H. G., Buta, R. J., Paurel,'G., \& Fouqu\'e, P. 1991, Third Reference Catalogue of Bright Galaxies
(Springer, New York).}

\hi{Wirth, D. 1996, Ph.\ D.\ Thesis, University of California at
Santa Cruz.}

\hi{Yee, H. K. C., Ellingson, E., \& Carlberg, R. C. 1996, \apjs 102, 269.}

\hi{Yoshii, Y. 1993, \apj 403, 552.}

\vfill
\clearpage


\begin{deluxetable}{lccccc}
\scriptsize
\tablecaption{Photometric Parameters for Some Recent
Faint Field Surveys\label{surveytab}}
\tablewidth{0pt} 
\tablehead{
\colhead{Survey}    		&       \colhead{Band}			&
\colhead{Seeing}      		&	\colhead{Magnitude Limits}	&
\multicolumn{2}{c}{Limiting Surface Brightness}	  
\\ \cline{5-6}
\colhead{}  	 		&       \colhead{} 			&
\colhead{(\arcsec)}      	&	\colhead{} 			&
\colhead{Photometry}   		&       \colhead{Detection}		
}
\startdata
CFRS\tablenotemark{a} 	& $I_{AB}$	&$0.6\arcsec-1\arcsec$	& $17.5<I_{AB}<22.5$		&28.0$\,I_{AB}/\Box\arcsec$	& $25.5\,I_{AB}/\Box\arcsec$	\nl
LCRS\tablenotemark{b} 	&Gunn-$r$	&$\sim1.3\arcsec$& $15\le r_{iso}\le17.7$		& $\sim23\,r/\Box\arcsec$	& $\sim22\,r/\Box\arcsec$ 	\nl
CNOC \tablenotemark{c}	&Gunn-$r$,$B_{AB}$& 	?	 & $r<20.5,21.5,22.0$			&\tablenotemark{j} 	& ?			\nl
AutoFib\tablenotemark{d,e}&$b_J$ 		&	?	& $19.5<b_J<22$			& 26.5$\,b_J/\Box\arcsec$	& $\sim25\,b_J/\Box\arcsec$	\nl
Norris\tablenotemark{f} & Gunn-$r$,$B_{AB}$&$\sim2.0\arcsec$& $r_{core}\le21.7,r_{tot}<20$	&\tablenotemark{k} 	& 24.1$\,r/\Box\arcsec$	\nl
Gardner\tablenotemark{g}& $K$\tablenotemark{i}	&	?	& $K\le15$			& \tablenotemark{l}	& 22.5$\,I$,18.5$\,K/\Box\arcsec$ \nl
Glazebrook\tablenotemark{h}& $K$	&$1.9\arcsec-3.5\arcsec$& $K\lta17$			& \tablenotemark{m}	& $\sim19\,K/\Box\arcsec$ \nl
\enddata
\tablenotetext{a}{Lilly et al.\ 1995a}
\tablenotetext{b}{Shectman et al.\ 1996}
\tablenotetext{c}{Yee et al.\ 1996, Lin et al.\ 1997}
\tablenotetext{d}{Ellis et al.\ 1996}
\tablenotetext{e}{Parameters refer only to AutoFib-faint sample; luminosity
function includes data from other surveys with different selection criteria}
\tablenotetext{f}{Small	et al.\ 1997}
\tablenotetext{g}{Gardner et al.\ 1996}
\tablenotetext{h}{Glazebrook et al.\ 1994}
\tablenotetext{i}{$K$ photometry performed only for galaxies detected in $I$}
\tablenotetext{j}{Variable aperture magnitudes used}
\tablenotetext{k}{FOCAS total magnitudes used}
\tablenotetext{l}{10$\arcsec$ apertures used}
\tablenotetext{m}{$4\arcsec-8\arcsec$ apertures used, depending on seeing.}
\end{deluxetable}
\vfill
\clearpage

\section{Figure Captions}

\figcaption[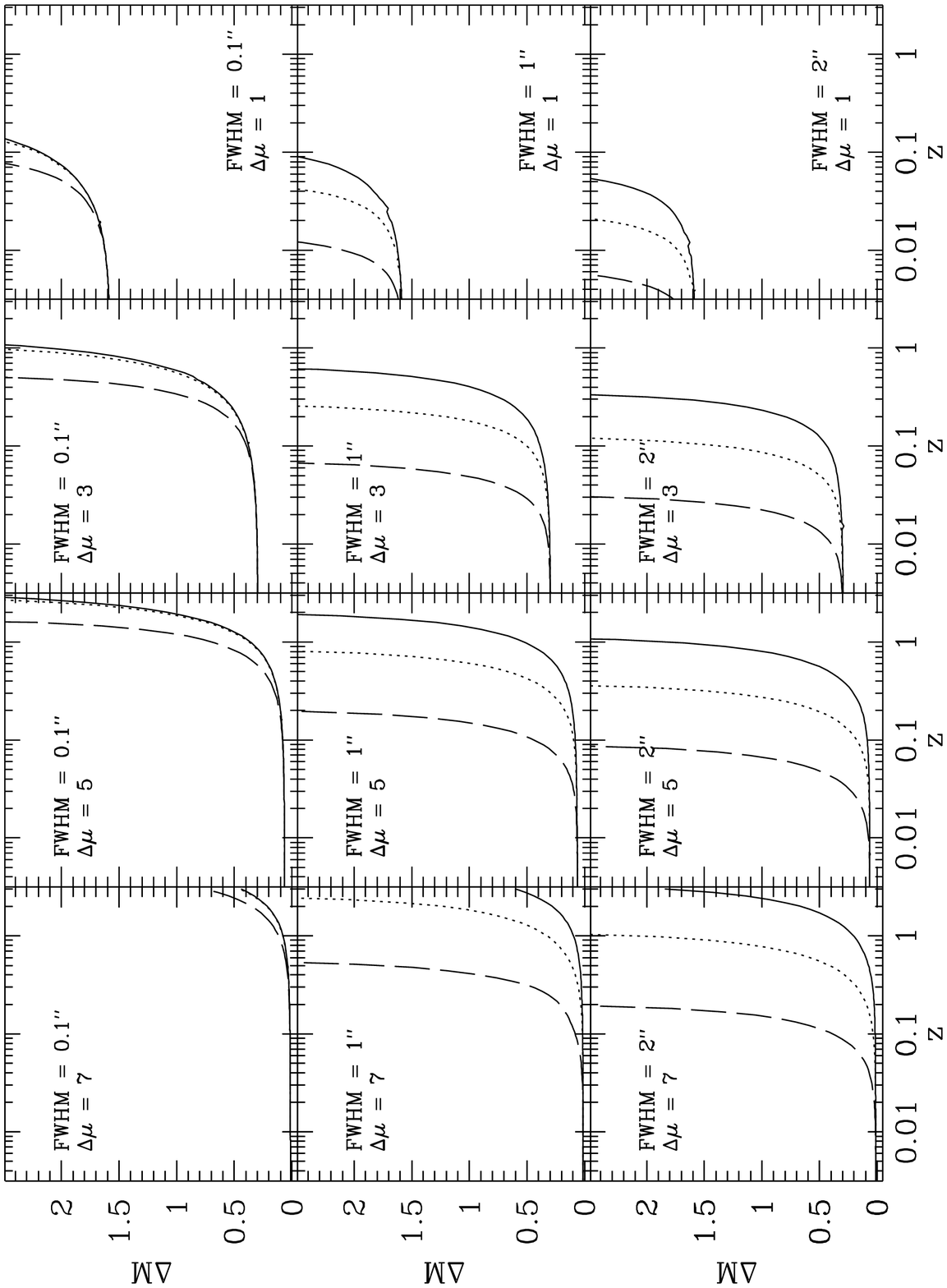,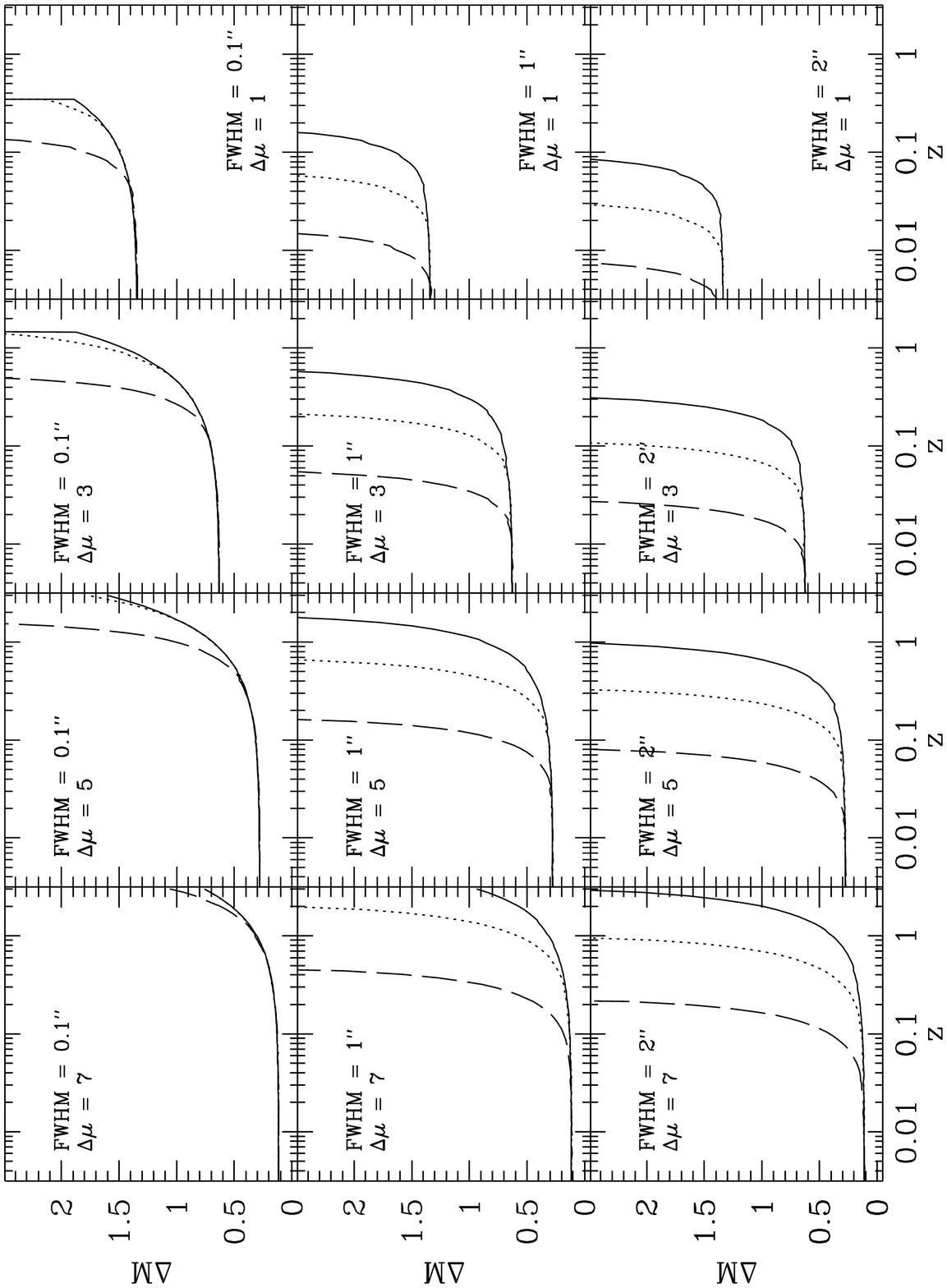] { [a] The fraction of the light detected
above a limiting isophote, represented as the error in the derived
absolute magnitude (i.e.\ $\Delta M(z) = -2.5\log{f(z)}$), is plotted
as a function of redshift, for different seeing (from top to bottom:
FWHM=0.1\arcsec, 1.0\arcsec, 2.0\arcsec, for a Moffat profile with
$\beta=5$), different galaxy surface brightnesses relative to the
limiting isophote (left to right:
$\Delta\mu=\mu_{lim}-\mu_0=7,5,3,1$), and different galaxy sizes
($\alpha=0.25\kpc,1\kpc,3\kpc$: dashed, dotted, and solid lines,
respectively), for circularly symmetric galaxies with exponential
surface brightness distributions.  $f(z)$ includes the effect of
cosmological dimming on reducing the fraction of light seen at high
redshift, but does not include the $k$-correction, which can modify
the effective central surface brightness at high redshift.  We have
assumed $H_0=50\hnot$ and $\Omega_0=1$.  For larger values of the
Hubble constant, the upturn in $\Delta M(z)$ will occur at larger
redshifts; however, the characteristic size of the galaxies in Figure
\ref{mualphafig} will become smaller.  [b] Same as [a], but for a de
Vaucouleurs profile.  See Figure \ref{petrosianfig} for a treatment of
aperture magnitudes and Petrosian magnitudes.
\label{ffig}}

\figcaption[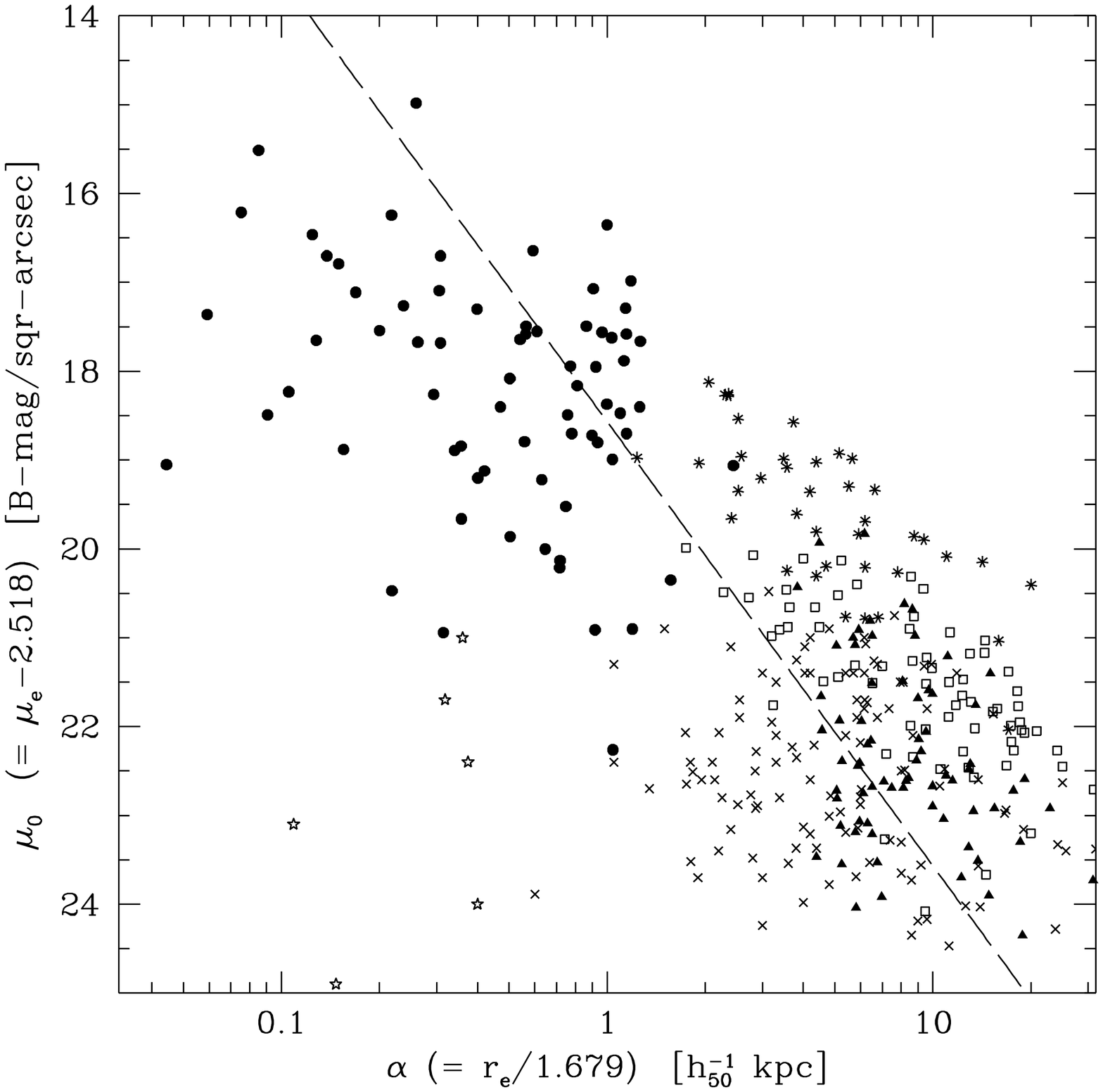] { The values of $\mu_0$ and $\alpha$ for a variety
of galaxy samples.  The solid circles and open squares are spiral
bulges and spiral disks, respectively, drawn from de Jong (1996a), and
the asterisks are ellipticals in A157 and A3574 from Jorgensen et al.\
(1995).  The solid triangles are low surface brightness galaxies
(LSBs) from the UGC catalog from Knezek (1993).  The crosses are
spiral disks from an assortment of other sources (Romanishin et al.\
1973, Boroson 1981, van der Kruit 1987, Sprayberry et al.\ 1995, de
Blok et al.\ 1995, McGaugh \& Bothun 1994) and are primarily selected
from the NGC, UGC, or POSS-II catalog of LSBs (Schombert et al.\
1992), all of which are either explicitly or implicitly angular
diameter limited field surveys.  The stars are the local group dwarf
spheroidals compiled by Caldwell et al.\ 1992.  The dashed line is the
locus of $M_B=-20$.  For the de Vaucouleurs profile galaxies (asterisks
and filled circles), the conversion from effective radius $r_e$ and
effective surface brightness $\mu_e$ to $\mu_0$ and $\alpha$ is as
given in the text.  All data has been converted to the Johnson $B$
magnitude system.  The data are chosen to represent the existing range
of galaxy morphologies and luminosities, and not to argue for a
particular relation between $\alpha$ and $\mu_0$ for a given profile
type.
\label{mualphafig}}

\figcaption[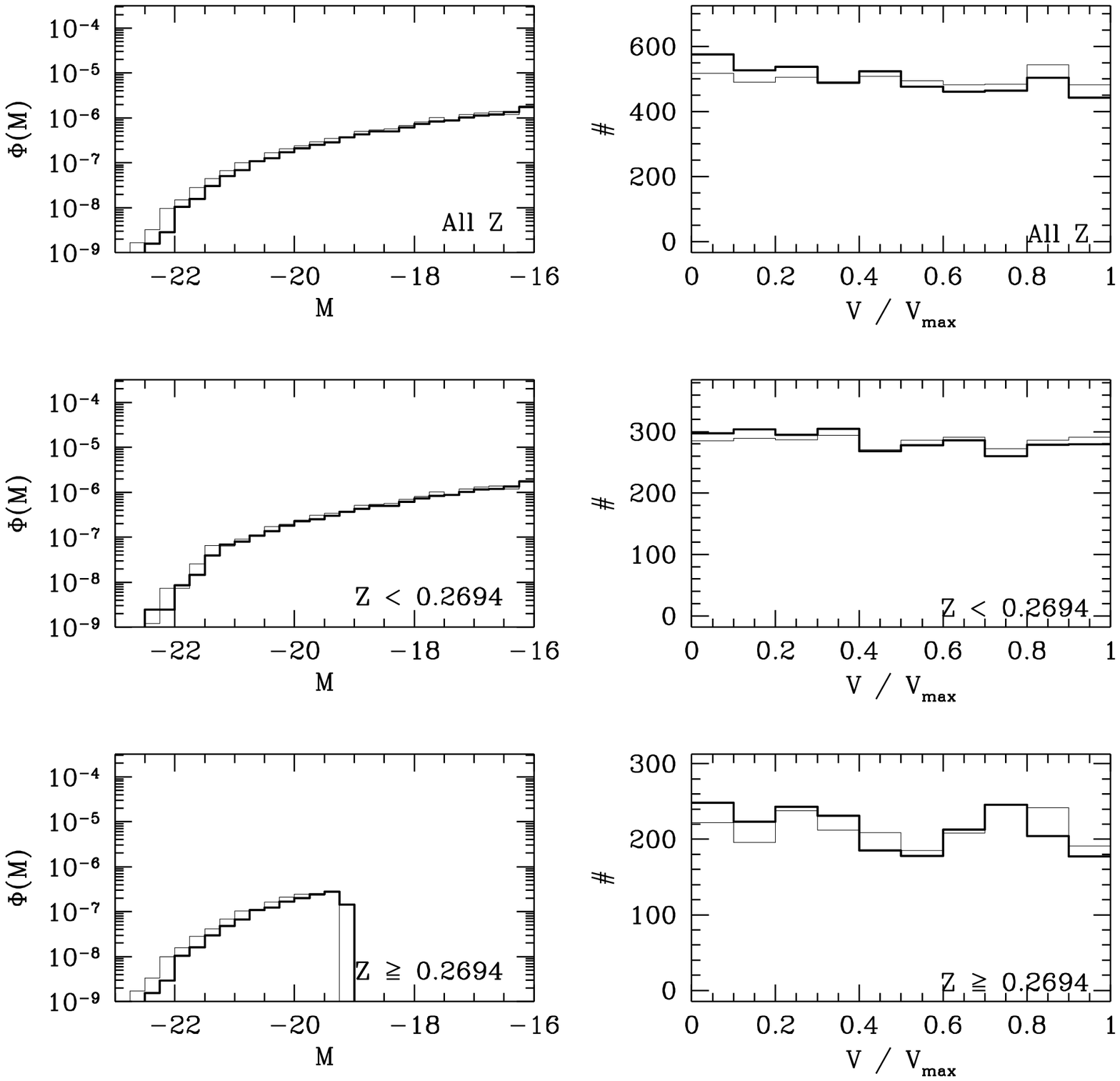,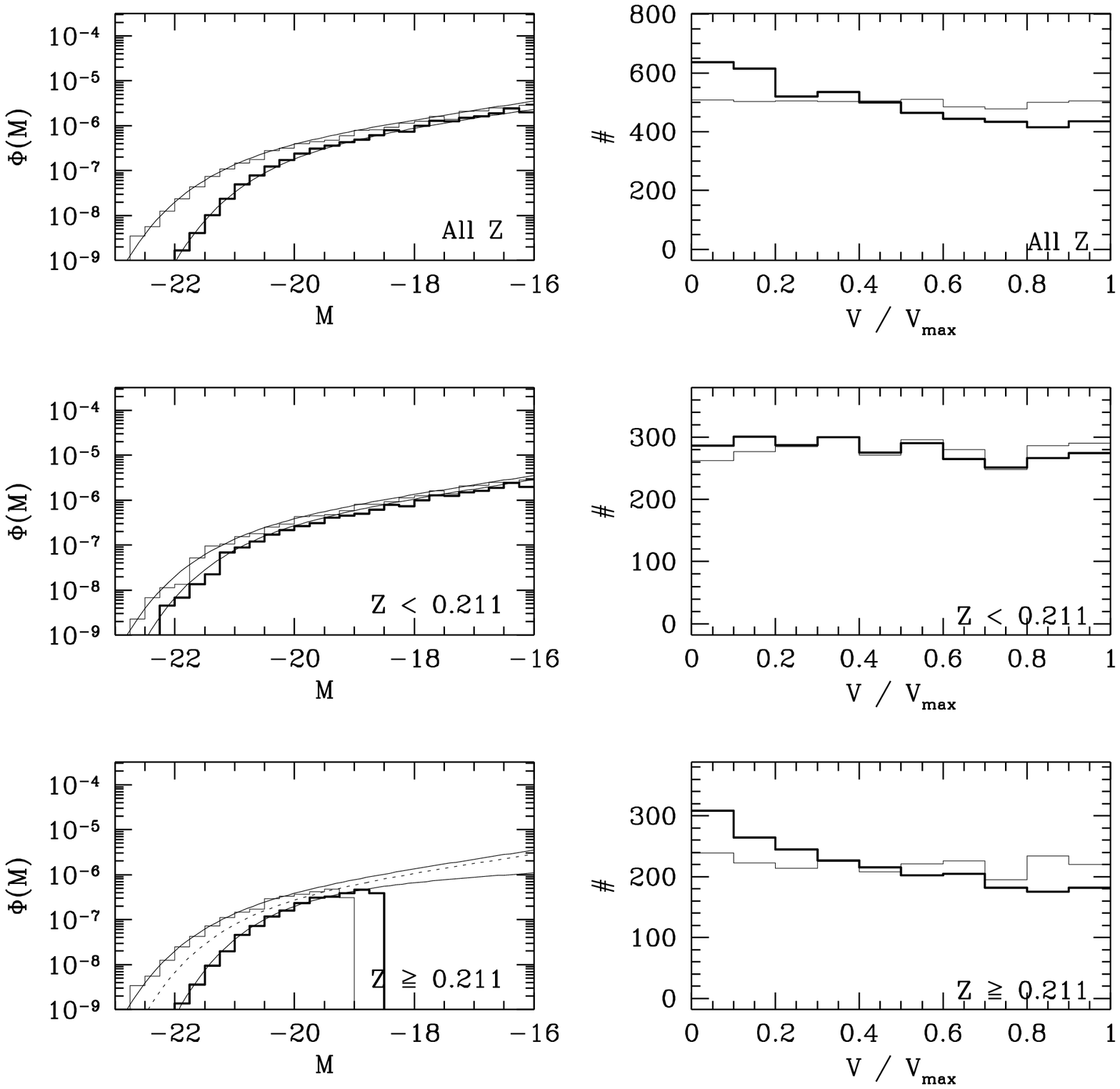,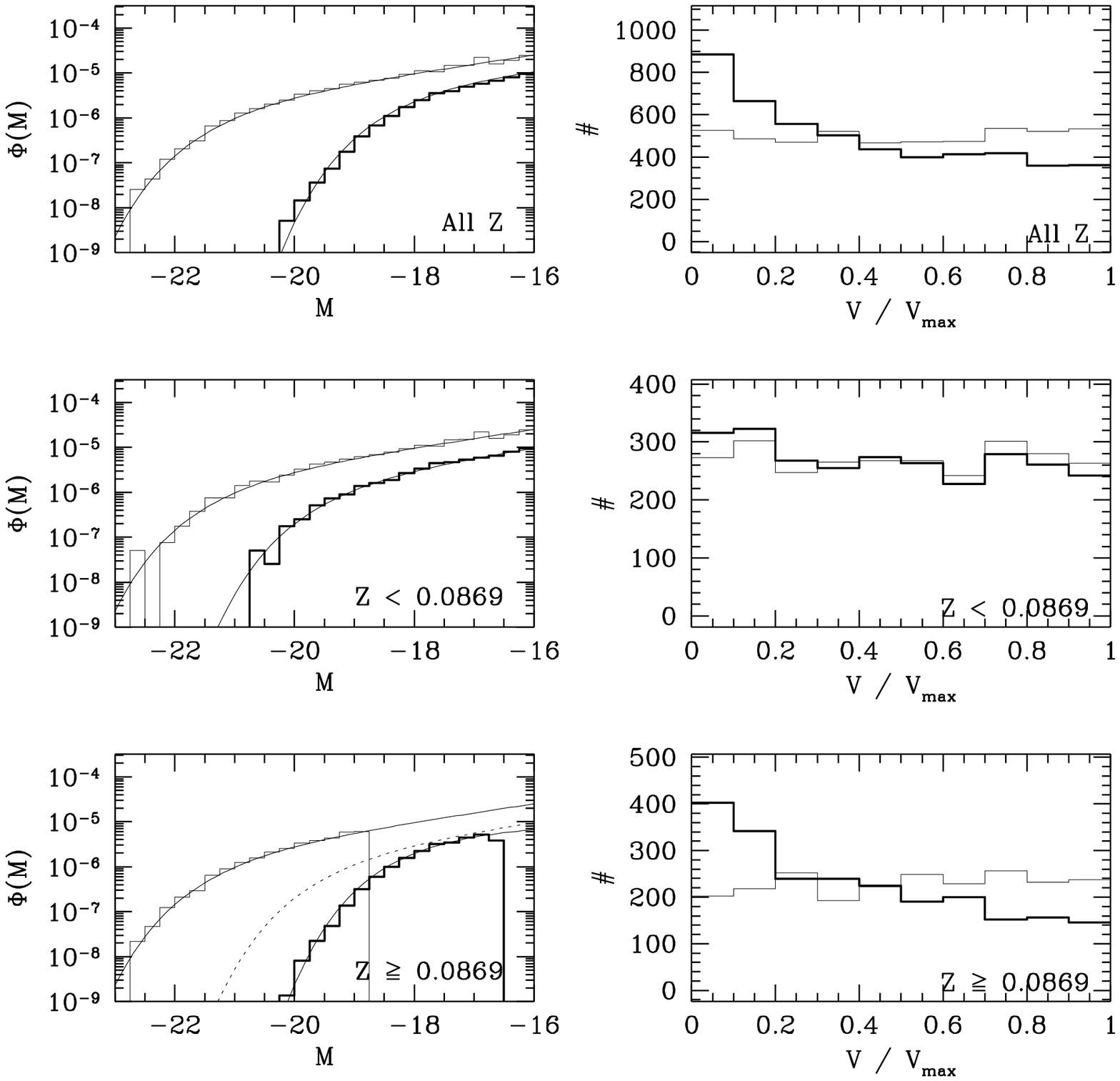,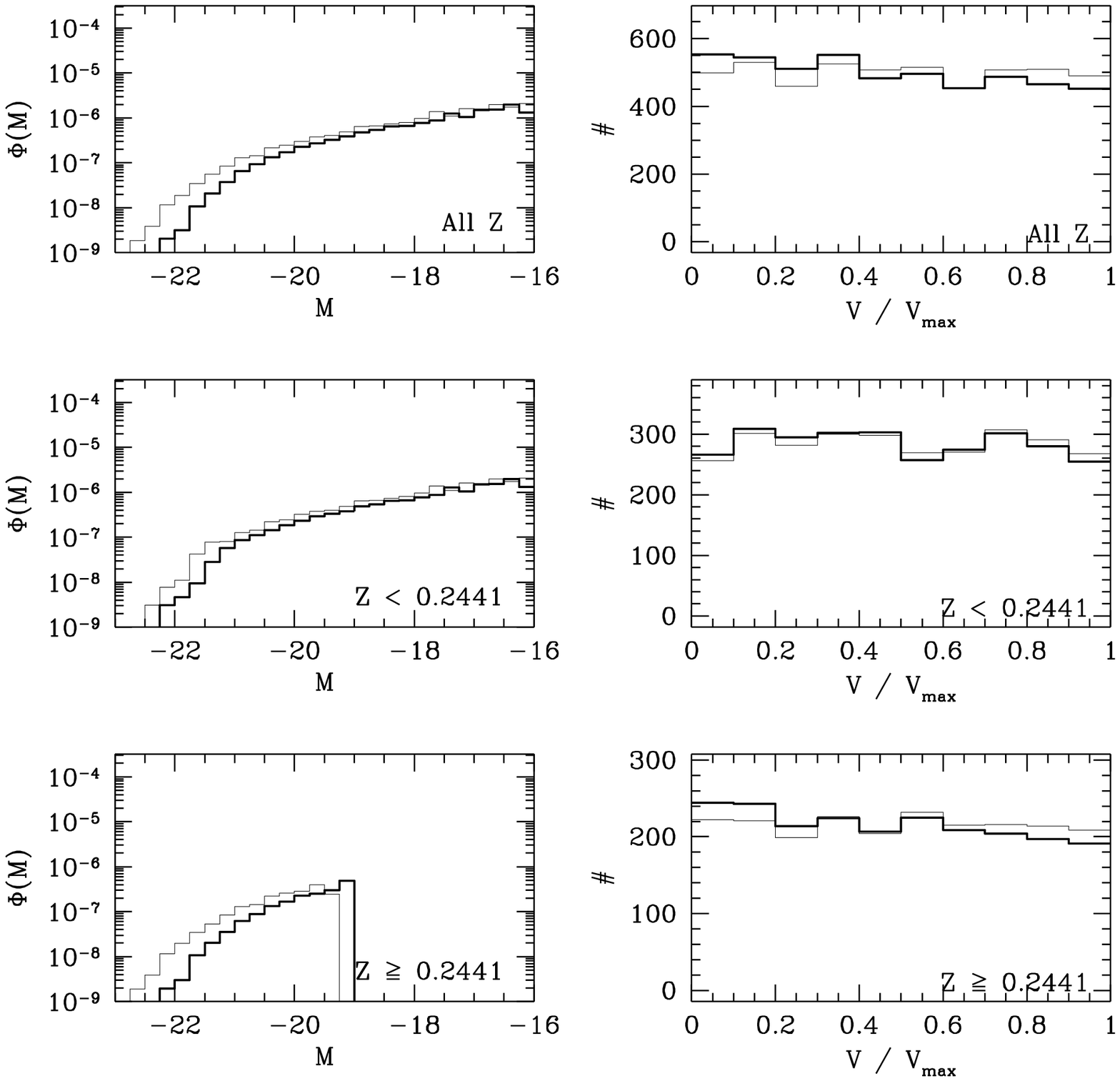,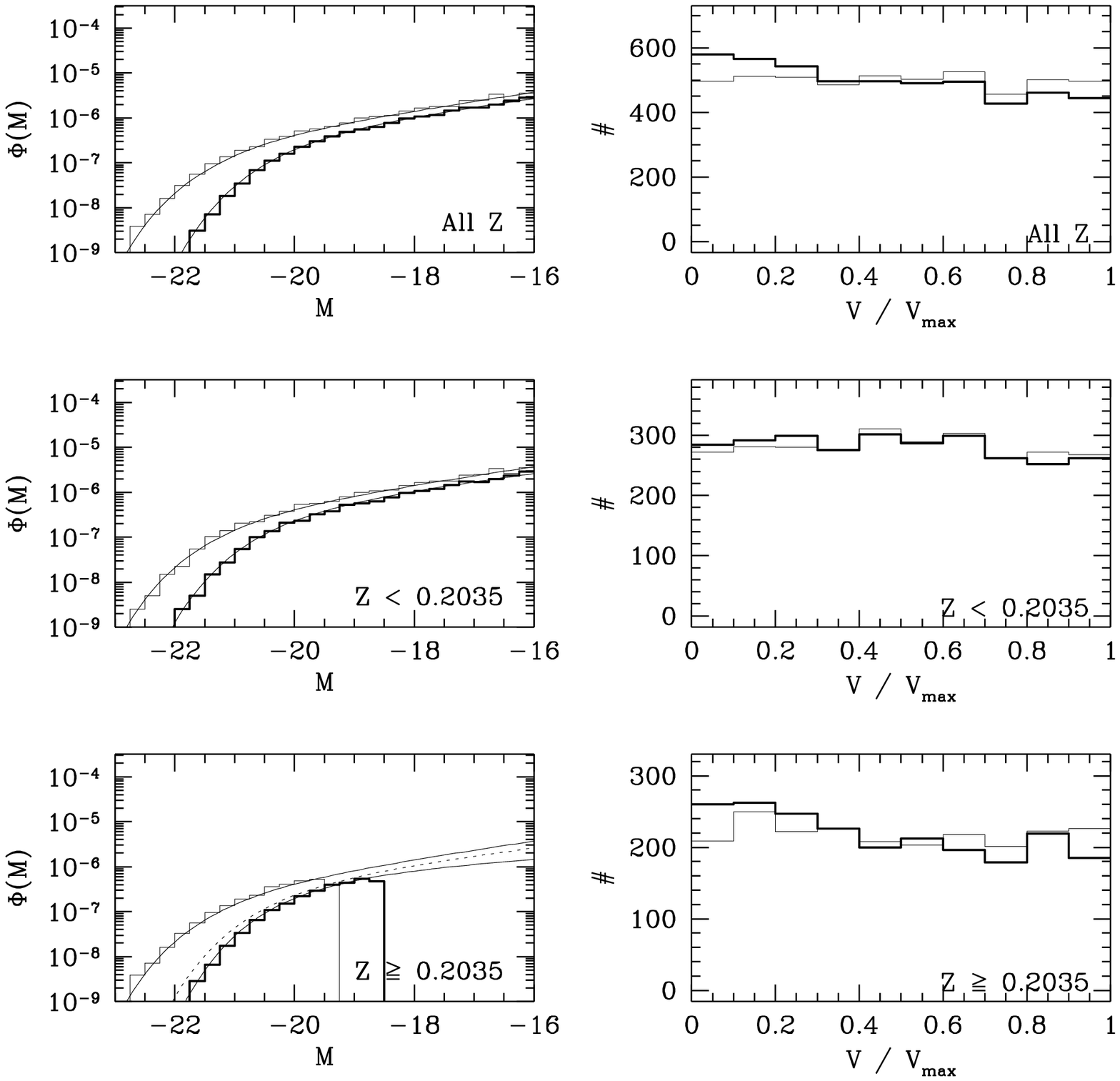,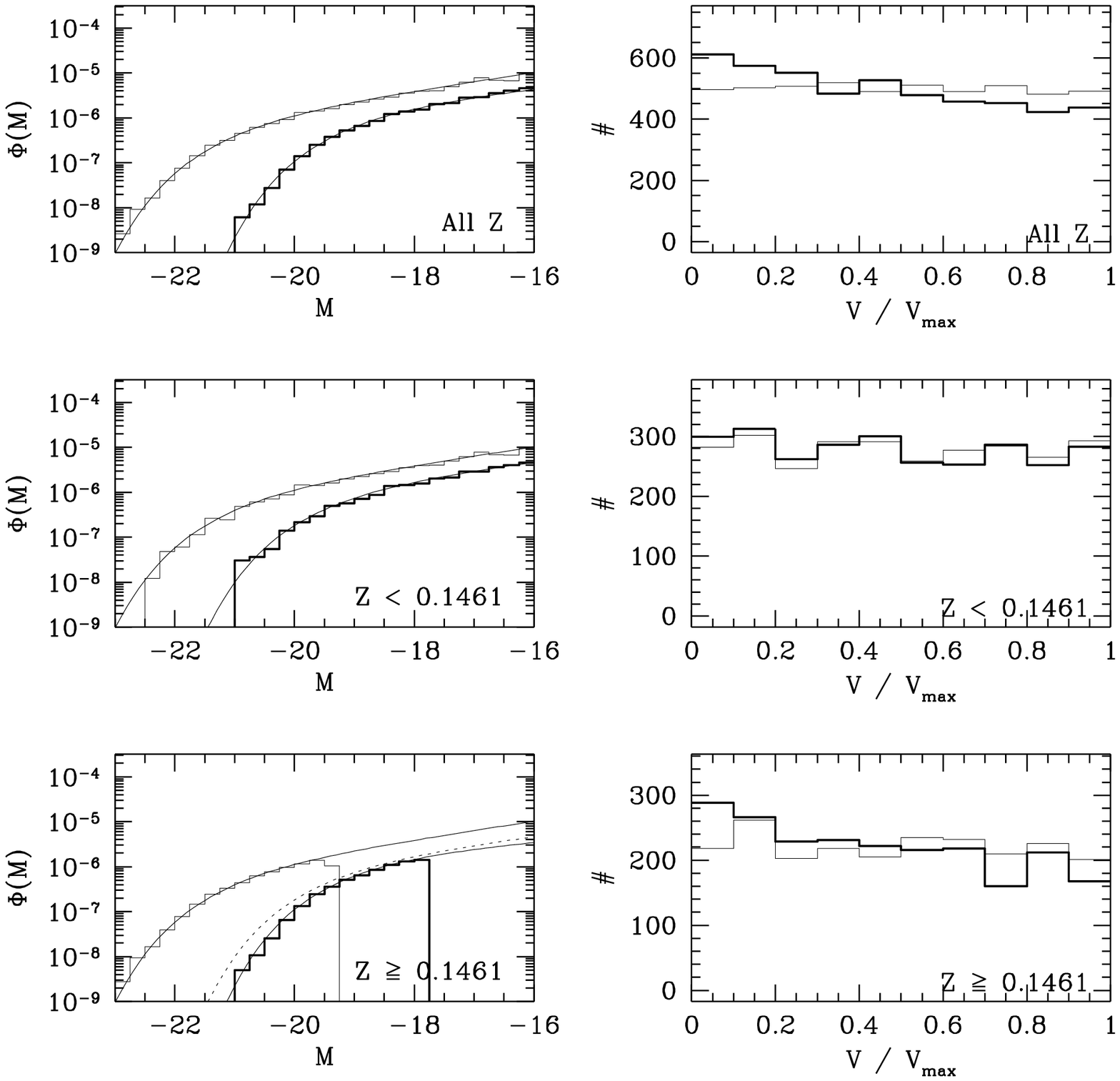] { The
luminosity functions (left column) and $V/V_{max}$ distributions
(right column) for samples of galaxies with different central surface
brightnesses [(a) $\mu_0=20\surfb$, (b) $\mu_0=22\surfb$, \& (c)
$\mu_0=24\surfb$ for exponential profile galaxies, and (d)
$\mu_0=20\surfb$, (e) $\mu_0=22\surfb$, \& (f) $\mu_0=24\surfb$ are
for de Vaucouleurs profile galaxies], calculated using standard
methods (heavy lines; eqn.\ \ref{Morigeqn}), and calculated using the
proper corrections for lost light (light lines; eqn.\ \ref{Meqn}).
The upper row is the luminosity function and $V/V_{max}$ distribution
derived from the entire sample.  The middle row is for the galaxies
which fall below the mean redshift of the sample, and the lower row is
for the galaxies falling above the mean redshift; the mean redshift
changes with surface brightness.  For the cases where the luminosity
function appears to evolves with redshift, Schechter luminosity
function fits are plotted superimposed to facilitate comparison; the
dotted line in the lower left panel is the luminosity function fit to
the low redshift data.  All samples have 5,000 galaxies, and assume a
limiting isophotal magnitude of $m_{lim}=25\surfb$, measured within an
outer isophote of $\mu_{lim}=22\surfb$, and $1\arcsec$ seeing, with a
$\beta=5$ Moffat profile distribution.  They do not include the
effects of random galaxy inclinations.  Note that the luminosity
function has a higher normalization for the lower surface brightness
galaxies, due to the systematically lower accessible volume, and thus
the smaller mean sample redshift (Figure \ref{ffig}; also Phillipps et
al.\ 1990 and references therein).
\label{lffig}}

\figcaption[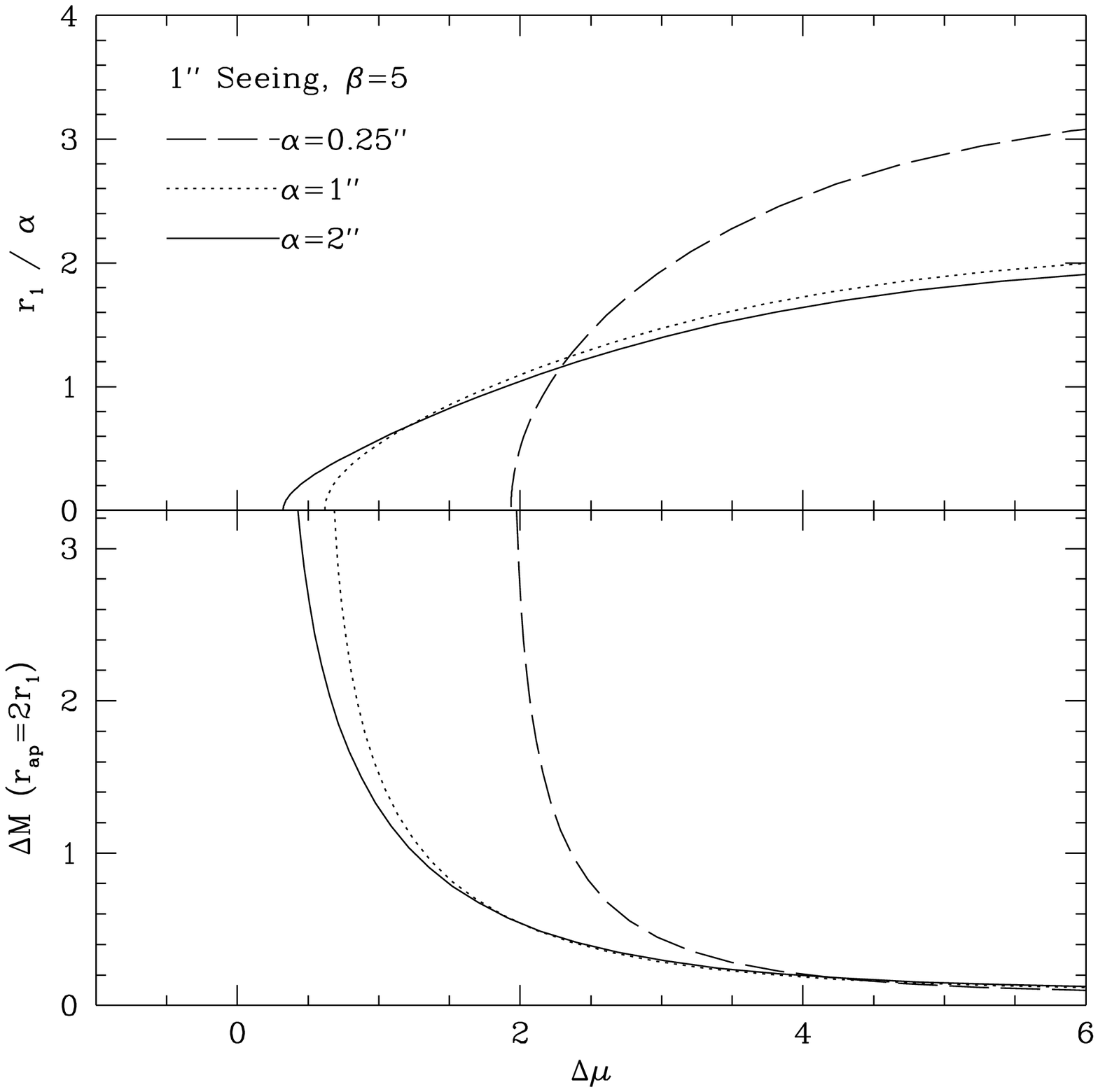,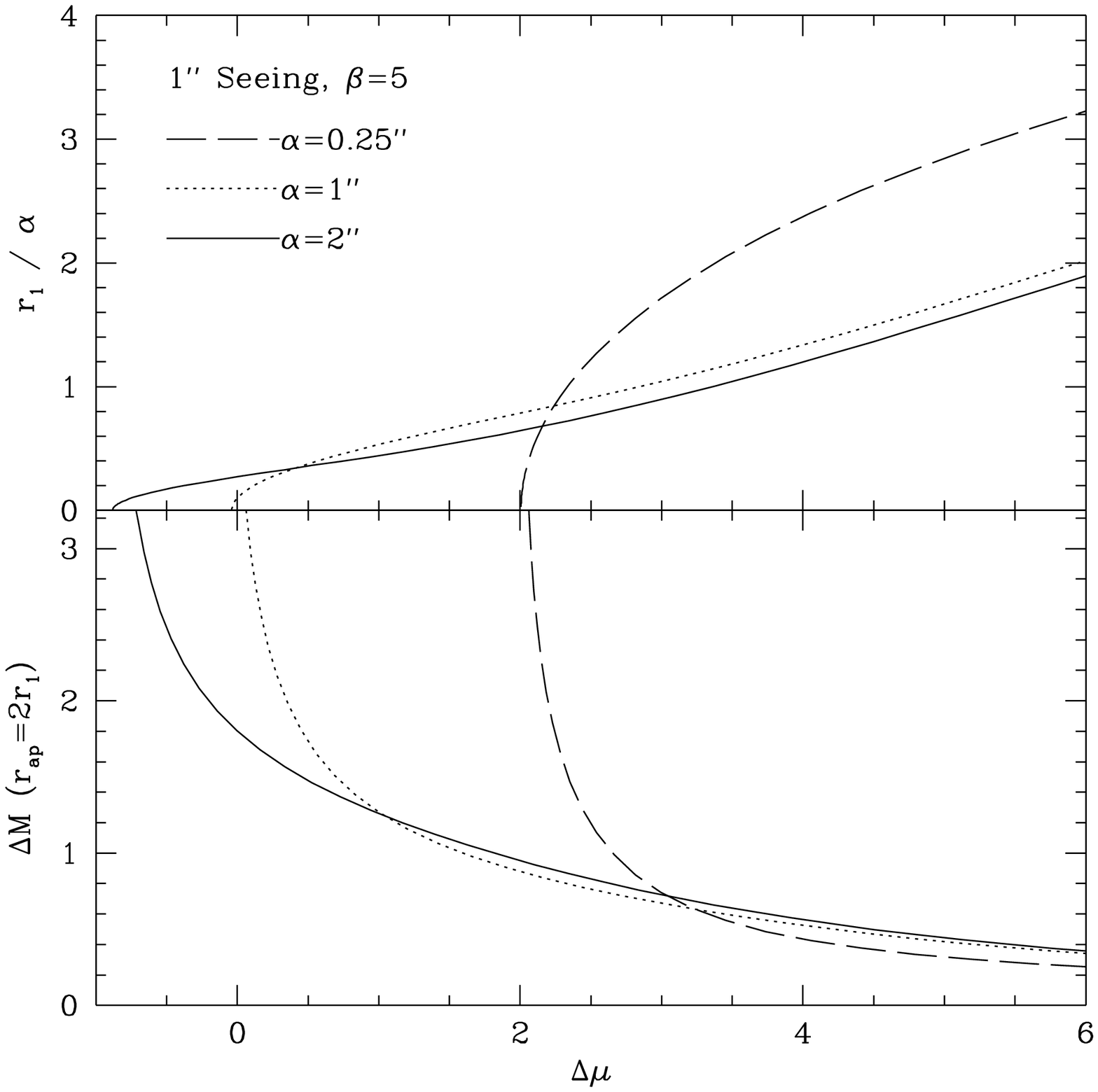] { The first moment radius $r_1$ measured
above some limiting surface brightness $\mu_{lim}$ (upper panel) and
the resulting error in the absolute Kron magnitude measured within
$2r_1$ (i.e.\ $\Delta M=-2.5\lg{f(2r_1)}$,lower panel), as a function
of $\Delta\mu=\mu_{lim}-\mu_0$, for different angular scale lengths
$\alpha=0.25\arcsec$, $1\arcsec$, \& $2\arcsec$ in the presence of a
Moffat PSF with $FWHM=1\arcsec$, for [a] exponential disks and [b] de
Vaucouleurs' profile ellipticals.  Lower surface brightness galaxies
(small $\Delta\mu$) have systematically underestimated Kron
magnitudes, because the value of $r_1$ is underestimated when only a
fraction of the galaxy's profile is used to calculate $r_1$.
Elliptical galaxies, which have a larger fraction of light at large
radii, are subject to larger errors in the measured luminosity than
exponential disks.  The curves are purely theoretical, and do not take
noise and other forms of measurement error into account.  For
reference, when measured within $r\rightarrow\infty$,
$r_1(\infty)/\alpha=2$ for a pure exponential profile, and
$r_1(\infty)/\alpha=3.8424$ for a pure de Vaucouleurs' profile.  The
biases are identical for the same for galaxies with the same ratio of
the $FWHM$ to the scale length $\alpha$.
\label{kronfig}}

\figcaption[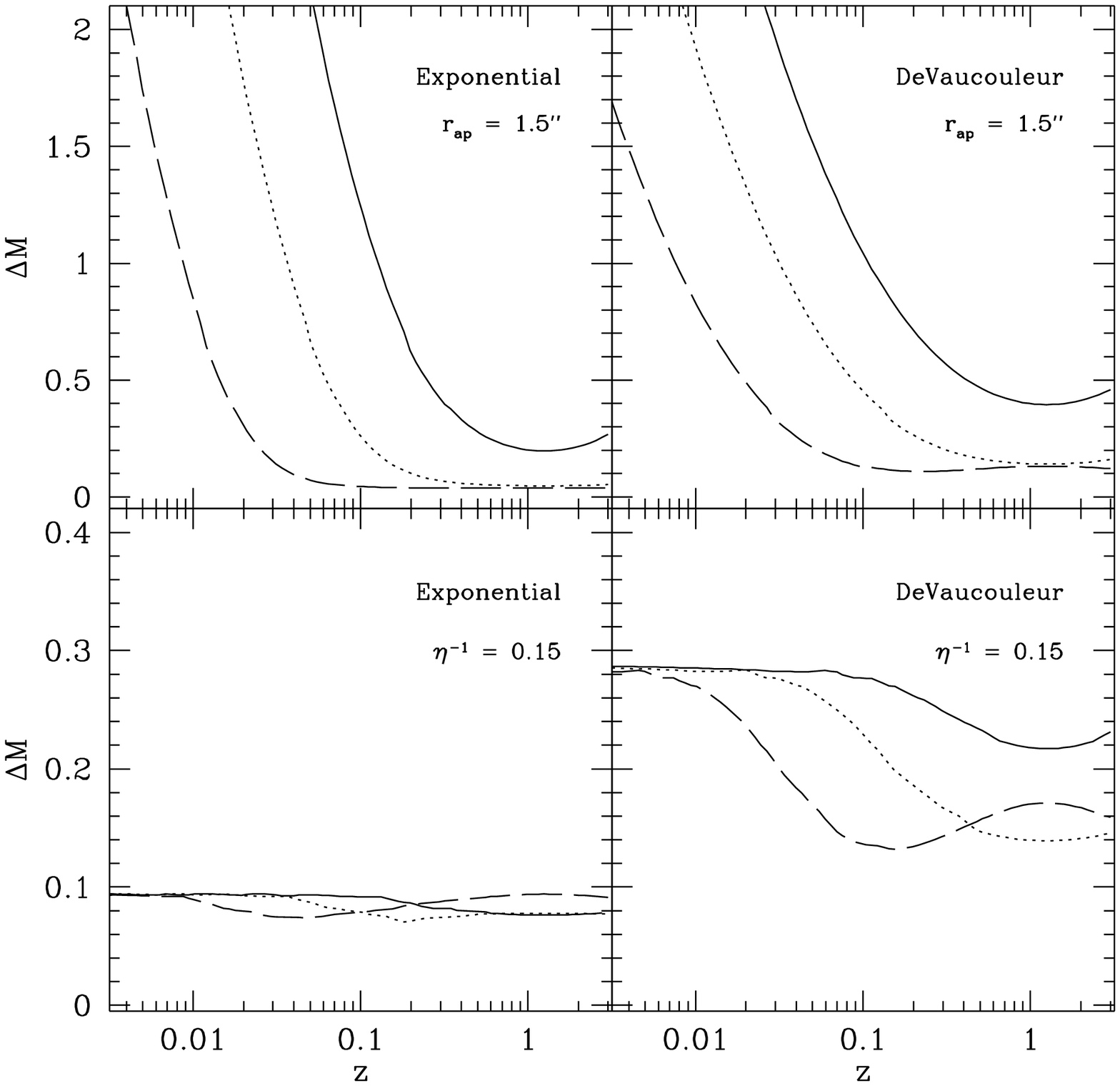] { The error in the measured absolute magnitude as a
function of redshift, for aperture and Petrosian magnitudes (top and
bottom, respectively), for both exponential (left) and de Vaucouleurs
(right) profile galaxies, in the presence of a $1\arcsec$ FWHM Moffat
profile PSF with $\beta=5$ (comparable to the middle row of Figure
\ref{ffig}).  As in Figure \ref{ffig}, the different lines represent
galaxies of different physical sizes ($\alpha=0.25\kpc,1\kpc,3\kpc$:
dashed, dotted, and solid lines, respectively).  Because both aperture
and Petrosian magnitudes scale linearly with surface brightness, the
given plots of $\Delta M$ are independent of galaxy surface
brightness.  Note the factor of 5 difference in scale between the
aperture magnitude plots and the Petrosian magnitude plots.
\label{petrosianfig}}

\vfill
\clearpage
\begin{figure}
\centerline{ \psfig{figure=f1a.ps,height=7.5in} }
\begin{flushright}{\bigskip\cap Figure \ref{ffig}[a]}\end{flushright}
\end{figure}
\vfill
\clearpage
\begin{figure}
\centerline{ \psfig{figure=f1b.ps,height=7.5in} }
\begin{flushright}{\bigskip\cap Figure \ref{ffig}[b]}\end{flushright}
\end{figure}
\vfill
\clearpage

\begin{figure}
\centerline{ \psfig{figure=f2.ps,height=7.5in} }
\begin{flushright}{\bigskip\cap Figure \ref{mualphafig}}\end{flushright}
\end{figure}
\vfill
\clearpage

\begin{figure}
\centerline{ \psfig{figure=f3a.ps,height=7.5in} }
\begin{flushright}{\bigskip\cap Figure \ref{lffig}[a]}\end{flushright}
\end{figure}
\vfill
\clearpage

\begin{figure}
\centerline{ \psfig{figure=f3b.ps,height=7.5in} }
\begin{flushright}{\bigskip\cap Figure \ref{lffig}[b]}\end{flushright}
\end{figure}
\vfill
\clearpage

\begin{figure}
\centerline{ \psfig{figure=f3c.ps,height=7.5in} }
\begin{flushright}{\bigskip\cap Figure \ref{lffig}[c]}\end{flushright}
\end{figure}
\vfill
\clearpage

\begin{figure}
\centerline{ \psfig{figure=f3d.ps,height=7.5in} }
\begin{flushright}{\bigskip\cap Figure \ref{lffig}[d]}\end{flushright}
\end{figure}
\vfill
\clearpage

\begin{figure}
\centerline{ \psfig{figure=f3e.ps,height=7.5in} }
\begin{flushright}{\bigskip\cap Figure \ref{lffig}[e]}\end{flushright}
\end{figure}
\vfill
\clearpage

\begin{figure}
\centerline{ \psfig{figure=f3f.ps,height=7.5in} }
\begin{flushright}{\bigskip\cap Figure \ref{lffig}[f]}\end{flushright}
\end{figure}
\vfill
\clearpage

\begin{figure}
\centerline{ \psfig{figure=f4a.ps,height=7.5in} }
\begin{flushright}{\bigskip\cap Figure \ref{kronfig}[a]}\end{flushright}
\end{figure}
\vfill
\clearpage

\begin{figure}
\centerline{ \psfig{figure=f4b.ps,height=7.5in} }
\begin{flushright}{\bigskip\cap Figure \ref{kronfig}[b]}\end{flushright}
\end{figure}
\vfill
\clearpage

\begin{figure}
\centerline{ \psfig{figure=f5.ps,height=7.5in} }
\begin{flushright}{\bigskip\cap Figure \ref{petrosianfig}}\end{flushright}
\end{figure}
\vfill
\clearpage

\end{document}